\documentclass[twocolumn,showpacs,preprintnumbers,amsmath,amssymb]{revtex4-1}

\usepackage{graphicx}
\usepackage{dcolumn}
\usepackage{bm}
\usepackage{multirow}
\usepackage{float}
\usepackage{pdfpages}

\def\done{$d_1$}
\def\dtwo{$d_2$}
\def\dthr{$d_3$}

\makeatletter
\providecommand \@ifxundefined [1]{%
 \@ifx{#1\undefined}
}%
\providecommand \@ifnum [1]{%
 \ifnum #1\expandafter \@firstoftwo
 \else \expandafter \@secondoftwo
 \fi
}%
\providecommand \@ifx [1]{%
 \ifx #1\expandafter \@firstoftwo
 \else \expandafter \@secondoftwo
 \fi
}%
\providecommand \natexlab [1]{#1}%
\providecommand \bibnamefont  [1]{#1}%
\providecommand \bibfnamefont [1]{#1}%
\providecommand \citenamefont [1]{#1}%
\providecommand \href@noop [0]{\@secondoftwo}%
\providecommand \href [0]{\begingroup \@sanitize@url \@href}%
\providecommand \@href[1]{\@@startlink{#1}\@@href}%
\providecommand \@@href[1]{\endgroup#1\@@endlink}%
\providecommand \@sanitize@url [0]{\catcode `\\12\catcode `\$12\catcode
  `\&12\catcode `\#12\catcode `\^12\catcode `\_12\catcode `\%12\relax}%
\providecommand \@@startlink[1]{}%
\providecommand \@@endlink[0]{}%
\providecommand \url  [0]{\begingroup\@sanitize@url \@url }%
\providecommand \@url [1]{\endgroup\@href {#1}{\urlprefix }}%
\providecommand \urlprefix  [0]{URL }%
\providecommand \doibase [0]{http://dx.doi.org/}%
\providecommand \selectlanguage [0]{\@gobble}%
\providecommand \bibinfo  [0]{\@secondoftwo}%
\providecommand \bibfield  [0]{\@secondoftwo}%
\providecommand \BibitemOpen [0]{}%
\providecommand \BibitemShut  [1]{\csname bibitem#1\endcsname}%
\let\auto@bib@innerbib\@empty

\begin{document}

\preprint{}

\title{Phonon anharmonicity and negative thermal expansion in SnSe} 


\author{Dipanshu Bansal,$^1$ Jiawang Hong,$^1$ Chen W. Li,$^1$ Andrew F. May,$^1$ Wallace Porter,$^1$ Michael Y. Hu$^2$, Douglas L. Abernathy,$^3$ and Olivier Delaire$^{1,4}$}
\email{bansald@ornl.gov}
\email{olivier.delaire@duke.edu}
\affiliation{
$^1$Materials Science and Technology Division, Oak Ridge National Laboratory, Oak Ridge, Tennessee 37831, USA \\
$^2$Advanced Photon Source, Argonne National Laboratory, Argonne, Illinois 60439, USA\\
$^3$Quantum Condensed Matter Division, Oak Ridge National Laboratory, Oak Ridge, Tennessee 37831, USA \\
$^4$Mechanical Engineering and Materials Science, Duke University, Durham, North Carolina 27708, USA
}

\textbf{Notice: This manuscript has been authored by
UT-Battelle, LLC under Contract No. DE-AC05-
00OR22725 with the U.S. Department of Energy. The
United States Government retains and the publisher, by
accepting the article for publication, acknowledges that
the United States Government retains a non-exclusive,
paid-up, irrevocable, world-wide license to publish or
reproduce the published form of this manuscript, or
allow others to do so, for United States Government
purposes. The Department of Energy will provide
public access to these results of federally sponsored research
in accordance with the DOE Public Access Plan
(http://energy.gov/downloads/doe-public-access-plan).\vspace{0.0in}}

\date{\today}

\begin{abstract}
The anharmonic phonon properties of SnSe in the Pnma phase were investigated with a combination of experiments and first-principles simulations. Using inelastic neutron scattering (INS) and nuclear resonant inelastic X-ray scattering (NRIXS), we have measured the phonon dispersions and density of states (DOS) and their temperature dependence, which revealed a strong, inhomogeneous shift and broadening of the spectrum on warming. First-principles simulations were performed to rationalize these measurements, and to explain the previously reported anisotropic thermal expansion, in particular the negative thermal expansion within the Sn-Se bilayers. Including the anisotropic strain dependence of the phonon free energy, in addition to the electronic ground state energy, is essential to reproduce the negative thermal expansion. From the phonon DOS obtained with INS and additional calorimetry measurements, we quantify the harmonic, dilational, and anharmonic components of the phonon entropy, heat capacity, and free energy. The origin of the anharmonic phonon thermodynamics is linked to the electronic structure. 

\end{abstract}

\maketitle


\section{Introduction}

Thermoelectric materials are of current interest for cost-effective, reliable power generation applications, either by utilizing natural sources or recovering man-made waste heat \cite{Snyder2008, Zebarjadi2012}. In order to increase their heat-to-electricity conversion efficiency, thermoelectric materials need to have a low thermal conductivity, while maintaining a high electrical conductivity. A detailed understanding of phonons is necessary to rationalize both the thermodynamics and the thermal transport properties in thermoelectrics. Phonons are the main contributors to the entropy and heat capacity \cite{Wallace1, Wallace2, Grimvall_book, Fultz2010}. In addition, phonons are the dominant heat carriers in semiconductors, and understanding deviations from harmonic lattice dynamics is necessary to account for the bulk lattice thermal conductivity, $\kappa_{\rm lat}$. From a fundamental standpoint, phonons are sensitive to the nature of chemical bonding \cite{Rhyee2009, Nielsen2013, Lee2014, Chen2015} and the temperature dependence of phonon spectral functions often provides important insights into the anharmonicity of the interatomic potential, directly affecting thermal resistivity through phonon-phonon scattering \cite{Delaire2011, Shiga_2012, Chen2015}. The coupling between phonons and electronic structure also has an important effect on phonon frequencies (and linewidths) \cite{Grimvall_book, Delaire_2011_PNAS, Delaire_2008_PRL, Delaire_2008_PRB, Bansal_2015}, and electronic instabilities can produce large anharmonicity, helping to suppress $\kappa_{\rm lat}$ and improve thermoelectric efficiency \cite{Chen2015, Lee2014, Jiawang-SnSe-arxiv2016}.

Tin-selenide was recently reported to achieve an outstanding thermoelectric figure-of-merit, with Na-doped samples maintaining a high power factor over a broad range of temperatures \cite{Zhao2014, Sassi2014, CLChen2014, Zhao2016}. At high temperature, SnSe undergoes a continuous structural phase transition between the low-symmetry Pnma (below $T_c \simeq 805$\,K) and the higher-symmery Cmcm phases (above $T_c$).  This phase transition has been the subject of a number of early crystallographic and thermomechanical experimental investigations \cite{Chatto1986, Adouby, Wiedemeier1979, Chatto1984, Wiedemeier1981, Feutelais1996, Sharma1986, Balde1981, Dembovskii1963, Zhdanova1961}. More recently, the ultra-low lattice thermal conductivity of SnSe has attracted great interest for thermoelectric applications, and has been studied both experimentally and through {\it ab-initio} simulations \cite{Zhao2014, Chen2015, Carrete2014, Sassi2014, Ding2015, Loa2015, Popuri2016, CLChen2014, Guo2015, Tobola2015, Wei2015, Sanchez2015, YLi2015, YMHan2015, ZHGe2015}. The coupling between lattice distortion and electronic structure was recently shown to be responsible for the large anharmonicity, which persists in a broad range of temperatures, and leads to the low thermal conductivity \cite{Chen2015,Jiawang-SnSe-arxiv2016}.

\begin{figure}
\includegraphics[width=0.4\textwidth]{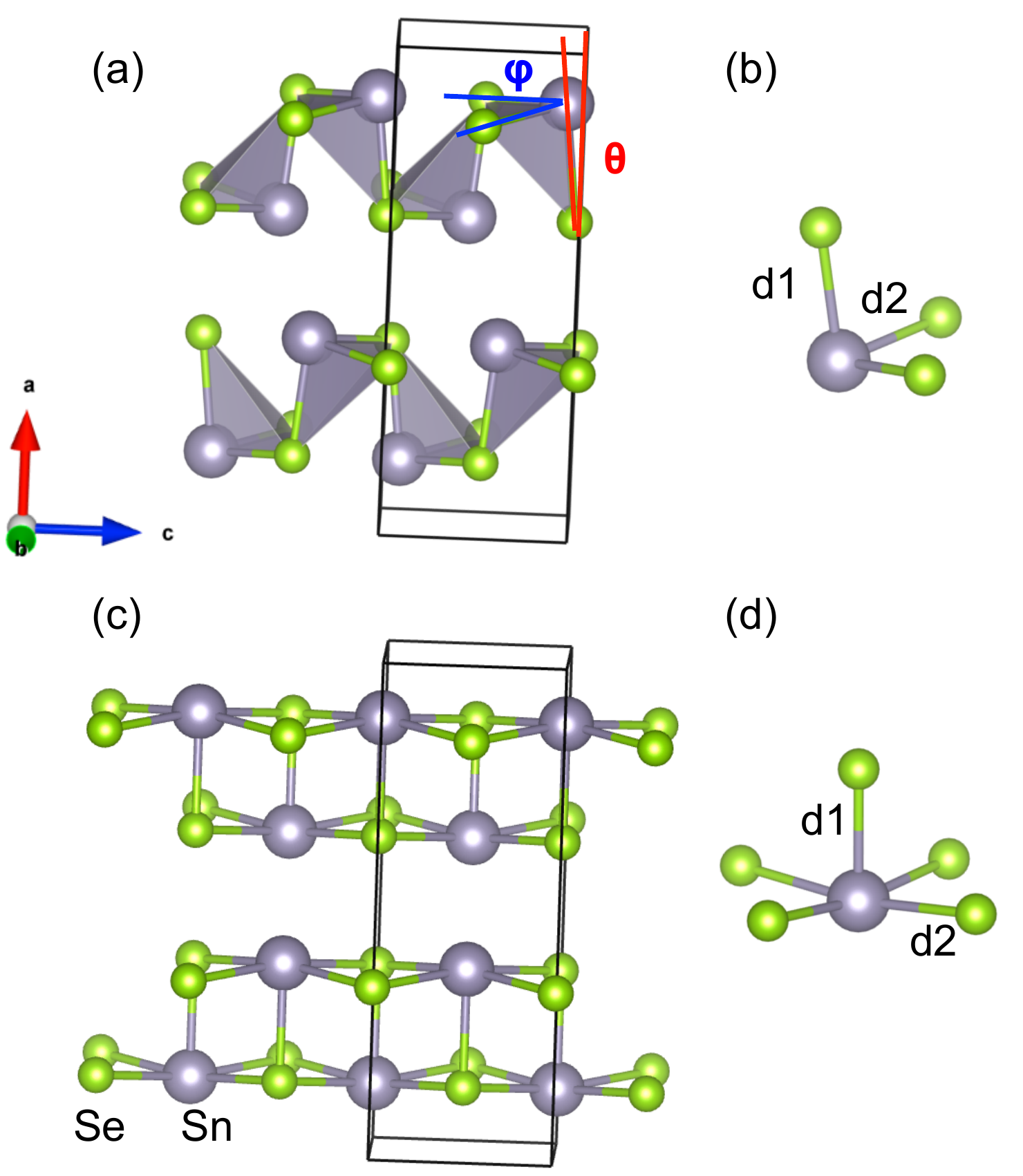}
\caption{\label{fig:rotation}
Crystal structure of SnSe in Pnma (a) and Cmcm (c) phases,
illustrating the double bilayer structure and Pnma distortion, and corresponding
Sn-Se bonding state (b,d). Sn atoms are in grey and Se atoms in green.
$d_1$ and $d_2$ are labels for bonds in out-of-plane and in-plane directions.  The orientation of crystal axes in the Cmcm structure is  
chosen to match the Pnma phase,
in order to facilitate the comparison. The rotation of pyramids (shaded in (a)) $\theta$ and the angle between 
two $d_2$ bonds $\phi$ in the Pnma structure are also shown in (a). The
conventional unit cell edges are indicated by black square box. }
\end{figure}

The crystal structure of SnSe is illustrated in Fig.~\ref{fig:rotation}. Importantly, x-ray and neutron diffraction studies showed that, in the Pnma phase, the $c$ axis, parallel to the direction of corrugation of Sn-Se bilayers below $T_c$, exhibits a negative thermal expansion (NTE) coefficient \cite{Adouby, Wiedemeier1979, Chatto1986}. The thermodynamic and atomistic origins of this effect in SnSe remain to be established. The NTE implies that  the thermodynamic generalized Gr\"{u}neisen tensor must have negative values for the diagonal element corresponding to the in-plane layer \cite{Grimvall_book}. Yet, this is in contrast to positive values predicted from quasiharmonic {\it ab-initio} simulations \cite{Zhao2014, Chen2015, Guo2015}. Further, our inelastic neutron scattering measurements showed pronounced differences in the behaviors of phonons propagating along in- and out-of-plane crystallographic axes \cite{Chen2015}. Since the coupling between the lattice distortion and the electronic structure is a strong source of anharmonicity \cite{Jiawang-SnSe-arxiv2016}, it is important to understand its role in the unusual thermal expansion of SnSe. 

Here, we investigate the origin of the NTE and anharmonic phonon thermodynamics in SnSe, focusing on the Pnma phase, using inelastic neutron scattering (INS) and nuclear resonant inelastic X-ray scattering (NRIXS) measurements, and first-principles simulations. Our measurements show a strong temperature dependence of the phonon frequencies, revealing a high degree of anharmonicity in the interatomic potential. We show how this anharmonicity is related to the structural phase transition. We model the thermal expansion and its anisotropic behavior, from first-principles, especially the lattice contraction parallel to the Sn-Se layers, and find good agreement with reported lattice parameter measurements. In addition, we quantify the harmonic, dilational, and anharmonic contributions of phonons to the entropy, internal energy and free energy.

\section{Sample Preparation}

Single crystals of SnSe were synthesized from high purity Sn and Se (Alfa Aesar, 99.999\%) in fused silica ampoules. After an initial reaction, a polycrystalline precursor was melted and subsequently crystallized while cooling at 0.4-0.5$^{\circ}$C/h; a 24\,hr annealing at $\sim$ 830$^{\circ}$C occurred before cooling to room temperature. One of the single crystals was ground into a fine powder for neutron scattering measurements of the phonon density of states.  Some properties of these samples were previously reported, and further details can be found in Ref.~\citenum{Chen2015}

\section{Spectroscopy}

\subsection{Inelastic Neutron Scattering}

We measured the phonon DOS of polycrystalline $\mathrm{SnSe}$ (mass $\sim8$\,g) using the time-of-flight wide angular-range chopper spectrometer (ARCS) at the Spallation Neutron Source (SNS) at Oak Ridge National Laboratory \cite{ARCS}. The powder sample was contained in a standard thin-walled aluminum can. Measurements at low ($5 \le T \le 300$\,K) and high temperature ($300 \le T \le 750$\,K) were performed using aluminum can inside a closed-cycle helium refrigerator and low-background resistive furnace. At low temperatures, we filled the sample chamber with low pressure of helium to facilitate cooling. We used two incident energies, $E_i=30$\,meV and 55\,meV, when combined it provided an entire phonon spectrum and  high resolution datasets to distinguish between closely spaced phonon peaks. An oscillating radial collimator was employed to minimize scattering from the sample environment. We followed the similar steps in data normalization, reduction to ${\bm Q}$-$E$ (momentum and energy transfer) space, background subtraction, multiphonon scattering, and removal of elastic peak as described in our previous work \cite{Bansal_2015}. 

The neutron scattering cross-sections are different for different elements. The phonon DOS derived from INS measurements is weighted by the respective scattering cross-sections ($\sigma$) and mass (m). The neutron-weighting factors for Sn, and Se are ${\sigma_{Sn}}/{m_{Sn}} = 0.0412$, and ${\sigma_{Se}}/{m_{Se}} = 0.1051$, respectively (in units of barns/amu). Due to the larger neutron-weighting factor for Se in comparison to Sn, phonon DOS is over-weighted by Se vibrational contributions. Accordingly, the neutron-weighted phonon DOS can be expressed as following:
\begin{eqnarray}
g_{NW} &=& \Big(\frac{\sigma_{Sn}}{m_{Sn}} g_{Sn}+ \frac{\sigma_{Se}}{m_{Se}} g_{Se}   \Big) \, /  \,\Big( \frac{\sigma_{Sn}}{m_{Sn}} + \frac{\sigma_{Se}}{m_{Se}} \Big) \, ,
\end{eqnarray}
\noindent where $g_{Sn}(E)$, and $g_{Se}(E)$ are the partial densities of states of Sn, and Se. For comparison of the experimental phonon DOS with the simulations, the neutron-weighting factor can be applied to the partial phonon DOS.

\subsection{Nuclear Resonant Inelastic X-ray Scattering}

NRIXS experiments on single-crystal SnSe platelets (natural Sn) were performed at Sector 30 beamline at the Advanced Photon Source, Argonne National Laboratory. Since SnSe exhibit anisotropic behavior, we could extract the direction-projected phonon DOS by changing the orientation of the sample with respect to the incident x-ray beam. For the $a$-axis projected phonon DOS, the sample was mounted such that the incident x-ray beam was parallel to $a$-axis, while for the projected phonon DOS perpendicular to $a$-axis, the incident beam was parallel to the b-c plane. The incident x-ray energy was 23.88\,keV, the nuclear resonance energy of $^{119}$Sn. The energy bandpass of incident beam was reduced to 1.2\,meV using a high resolution crystal monochromator \cite{Toellner2011}. The inelastic signal collected is comprised mainly of 23.88 keV nuclear fluorescence signal, delayed in time relative to the prompt pulse by 20-130\,nsec. The inelastic spectra were collected over an energy range of 160\,meV with 0.25\,meV steps. Repeat scans were taken and data were subsequently added. The $^{119}$Sn specific phonon DOS was extracted from the NRIXS data using the PHOENIX package \cite{Sturhahn2000}. The samples were single-crystal SnSe platelets (no isotopic enrichment), cut from INS crystals. The platelet samples were mounted with either transmission geometry (beam along $a$--axis) or grazing geometry (beam nearly parallel to $b-c$ plane). Because of the crystalline anisotropy, the direction-projected partial DOS for each configuration provides information about different Sn motions ($||a$ or $\perp a$, respectively).

\section{Modeling}

\subsection{Structural Relaxation and Phonon Dispersions}\label{DFT_simulations}

First-principles simulations were performed in the framework of density functional theory (DFT) as implemented in the Vienna Ab initio Simulation Package (VASP 5.3) \cite{Kresse1993, Kresse1996b, Kresse1996}.  A $6\times12\times12$ Monkhorst-Pack electronic \emph{k}-point mesh was used for all of our simulations, with a plane-wave cut-off energy of 500\,eV. The projector-augmented-wave potentials explicitly included four valence electrons for Sn ($5s^25p^2$), and 6 for Se ($4s^24p^4$). The lattice parameters and atomic positions were optimized until forces on all atoms were smaller than 1\,meV\,\AA$^{-1}$. We carefully evaluated the accuracy of our phonon calculations, comparing the local-density approximation (LDA) exchange-correlation (XC) functional \cite{refLDA} and the generalized gradient approximation (GGA) with XC functional given by the Perdew-Burke-Ernzerhof (PBE) parametrization \cite{refPBE}. The relaxed unit cell parameters are compared with the experimentally reported structure (Ref.~\citenum{Adouby}) at 300\,K in Table~\ref{lattice_parameters_SnSe}. The agreement between the LDA and PBE structures and the experimental structure is good, with, as expected, LDA underestimating the lattice constants and PBE overestimating them, by opposite amounts.

\begin{figure}
\begin{center}
\includegraphics[trim=0cm 0.0cm 0cm 0.10cm, clip=true, width=0.45\textwidth]{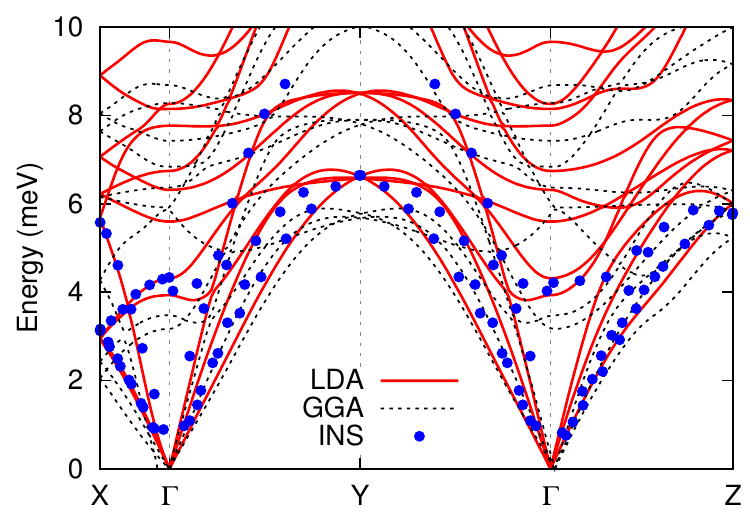}
\end{center}
\caption{\label{Phonon_dispersion_LDA_GGA} Phonon dispersions of SnSe from DFT simulations with either LDA or  GGA  exchange-correlation functionals (relaxed unit cells), compared with experimental INS data at 300\,K.}
\end{figure}

The phonon dispersions were calculated in the harmonic approximation, using the finite displacement approach as implemented in Phonopy \cite{Phonopy}, with the atomic forces in the distorted supercells obtained with VASP. To construct the force-constant tensor using the finite displacement approach in Phonopy \cite{Phonopy}, we computed 8 independent atomic displacements, each with atomic displacement amplitudes of one hundredth of lattice constants. The LDA phonon calculations used a $2\times 4\times 4$ supercell of the primitive cell, containing 256 atoms, while GGA calculations used up to $3\times 5\times 5$ supercells containing 600 atoms. We compare the phonon dispersion and phonon DOS of SnSe at 300\,K with experimental phonon dispersions measured with INS in Fig.~\ref{Phonon_dispersion_LDA_GGA}, and with the NRIXS and INS phonon DOS in Fig.~\ref{partial_DOS_SnSe} and~\ref{DOS_SnSe}, respectively. The experimental phonon dispersions in Fig.~\ref{Phonon_dispersion_LDA_GGA} were extracted from INS datasets originally reported by our group in Ref.~\citenum{Chen2015}. From these comparisons with INS measurements, one can see how the PBE simulations systematically underestimate the phonon frequencies in this material, while LDA simulations are much more accurate. Even with the larger supercell, the GGA phonon frequencies show a worse agreement with INS data than the LDA calculations. The mean phonon energy calculated from experimental phonon DOS at 300\,K (corrected for neutron weighting), GGA simulations, and LDA simulations are 12.92, 12.15, and 12.90\,meV, respectively. Additionally, the GGA phonon group velocities $v_{q,j}$ are on average $\sim$10\% lower than the LDA values near the $\Gamma$ point. Thus, first-principles calculations of the lattice thermal conductivity ($\kappa_{\rm lat}$) using GGA (Refs.~\citenum{Carrete2014,Ding2015}) are expected to carry a sizeable systematic error, since $\kappa_{\rm lat}$ is proportional to $v_{q,j}^2$, with a further dependence of phonon scattering rates on the dispersions, while LDA calculations are expected be more accurate \cite{Guo2015}.

\begin{table}
  \caption{Lattice constants and internal atomic coordinates for Pnma phase from DFT simulations compared with experimental data at 300\,K. Internal coordinates are normalized to the lattice constants. The number in parentheses are percentage deviation from the experimental data reported in Ref.~\citenum{Adouby}. Details are provided in the text.}
  \label{lattice_parameters_SnSe}
\begin{center}
  \begin{tabular}{c|c|c|c}
  \hline
  & LDA & GGA & Exp. \cite{Adouby} \\
  \hline
  a  (\AA) & 11.309 (1.68) &  11.756 (-2.20) & 11.502 \\
 b (\AA) & 4.119 (0.82) & 4.205 (-1.25) & 4.153 \\
 c (\AA) & 4.300 (3.37) & 4.547 (-2.18) & 4.450  \\   
 V (\AA$^3$) & 200.302 (5.77) &  224.776 (-5.74) & 212.567\\                     
\hline
  \end{tabular}
  \end{center}

  \begin{center}
  \begin{tabular}{c|cc|cc|cc}
  \hline
   & \multicolumn{2}{c|}{LDA} & \multicolumn{2}{c|}{GGA} & \multicolumn{2}{c}{Exp. \cite{Adouby}} \\
  \hline
   & x & z & x & z & x & z \\
 Sn  & 0.1169 &  0.0908 & 0.1206 & 0.1144 & 0.1208 & 0.1060 \\
Se  & 0.8580 &  0.4753 & 0.8551 & 0.4754 & 0.8551 & 0.4760 \\           
\hline
  \end{tabular}
  \end{center}
  
\end{table}

\subsection{Negative Thermal Expansion}
\begin{figure}
\begin{center}
\includegraphics[trim=5cm 15.0cm 27cm 0.0cm, clip=true, width=0.9\textwidth]{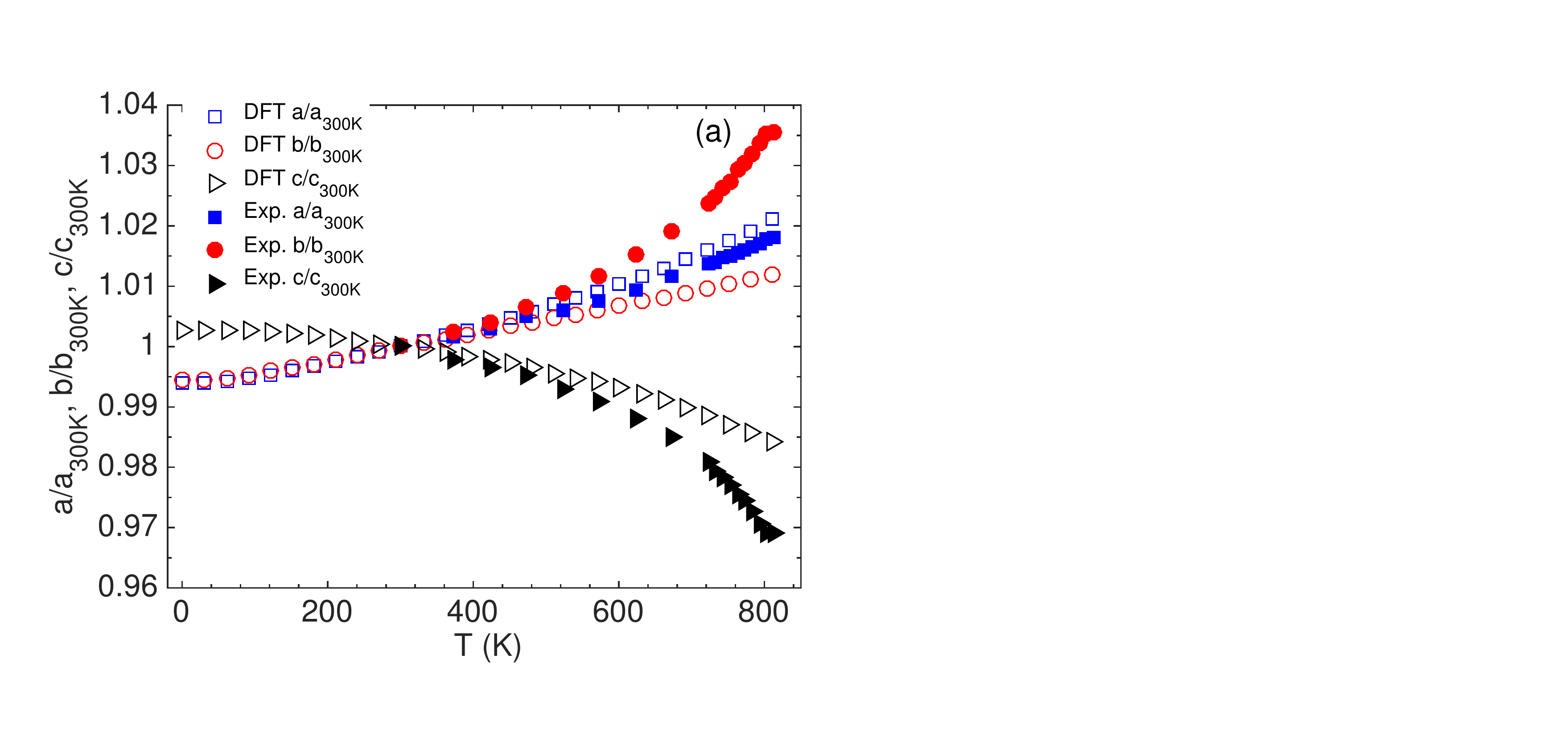}
\includegraphics[trim=34.5cm 0.0cm 4.7cm 2.0cm, clip=true, width=0.4\textwidth]{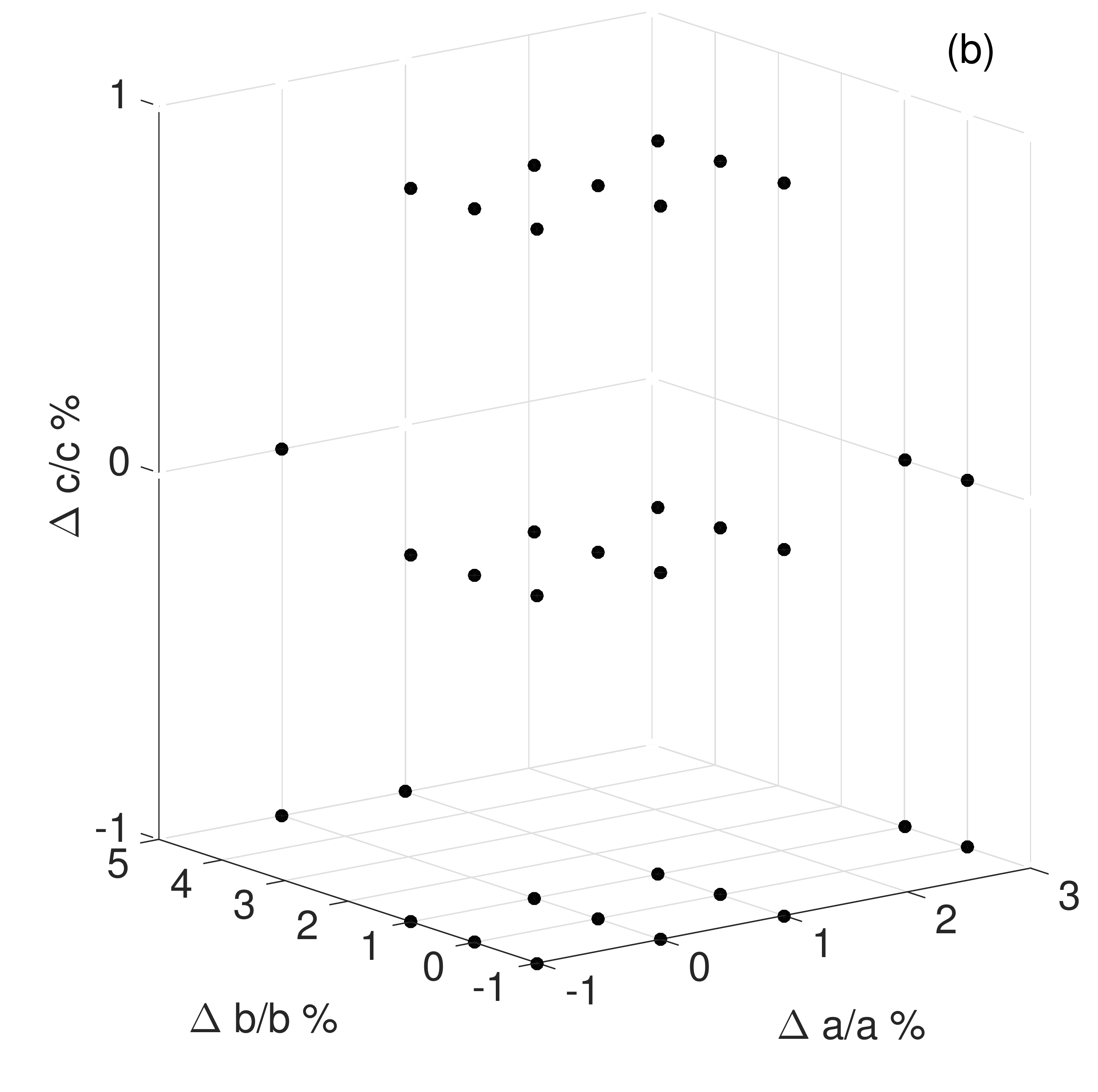}
\end{center}
\caption{\label{ABC_harmonic} (a) Lattice parameters calculated (open markers) from minimization of the free energy surface at different temperatures, compared with experimental data (solid markers) from Adouby \emph{et al.}~\cite{Adouby}. (b) Grid of isothermal strains computed with DFT to sample the free-energy surface.
}
\end{figure}

Non-cubic materials often exhibit anisotropic thermal expansion coefficients, reflecting the anisotropy in crystal structure and bonding. In the case of orthorombic SnSe, the anisotropy is particularly striking, with negative thermal expansion (NTE) of the $c$ axis, parallel to the direction of corrugation of Sn-Se layers (also corresponding to the direction of atomic motions across the phase transition), while the other in-plane axis ($b$) and the out-of-plane $a$--axis show normal, positive expansion \cite{Wiedemeier1979, Chatto1986, Adouby}. The relative change of lattice parameters with temperature reported by Adouby {\it et al.} in Ref.~\citenum{Adouby} are shown in Fig.~\ref{ABC_harmonic}. This distortion can also be analyzed as a rotation of SnSe$_3$ tetrahedra, as illustrated in Fig.~\ref{fig:rotation} and further discussed below. We proceed to rationalize this behavior based on first-principles thermodynamics and a microscopic analysis of bonding.

We start by noting that simple Gr\"{u}neisen parameter calculations are insufficient to explain the anisotropy and NTE along one or more crystallographic axes. Instead, one needs to compute the 3-D total free energy surface, whose temperature dependence is dominated by the phonon entropy. In order to clearly make this point, we start by performing the computation of Gr\"{u}neisen parameters. A generalized Gr\"{u}neisen tensor $\gamma_{ij}^{p}({\bm q})$ for a given polarization index $p$ and wave vector ${\bm q}$ is given by: \cite{Brugger1965,Cantrell1980} 
\begin{align}
\gamma_{ij}^{p}({\bm q}) = -\left\{\frac{1}{\omega^p({\bm q})}\left[\frac{\partial\omega^p({\bm q})}{\partial\eta_{ij}}\right]_T\right\}_{\eta=0},
\end{align}
where, $\eta_{ij}$ is the $(i,j)$ component of applied isothermal strain, and $\omega^p({\bm q})$ is the atomic vibration frequency. Summation over polarization index and wave vector, weighted by heat capacity at constant configuration $C_{\eta}$, yields an average value of a generalized Gr\"{u}neisen tensor. We have calculated the generalized Gr\"{u}neisen tensor from our DFT simulations of phonons $\gamma_{ij, {\rm DFT}}^{\rm avg}$. In the case of ${\rm SnSe}$, considering the symmetry of the second order tensor for the orthorhombic unit cell, non-diagonal components are exactly zero. Thus, the only isothermal strains required are $\eta_{11}$, $\eta_{22}$, and $\eta_{33}$, corresponding to stretching or compressing the lattice along crystallographic axes $a$, $b$, and $c$, respectively. Subsequently, the generalized Gr\"{u}neisen tensor, $\gamma_{ij, {\rm DFT}}^{\rm avg}$ is obtained by averaging over all polarizations and wave vectors, with each mode weighted by its heat capacity at constant configuration as follows:
\begin{align}\label{NTE1}
\gamma_{ij, {\rm DFT}}^{\rm avg} &= \begin{bmatrix}     1.44     &    0   &      0 \\
         0  &  1.35    &     0 \\
         0    &     0  & 0.64
         \end{bmatrix}.
\end{align}

 We have also calculated the  Gr\"{u}neisen tensor along high symmetry directions by averaging over polarizations only. Results are listed in Table~\ref{Anisotropic_Gruneisen_SnSe}. As can be seen from table~\ref{Anisotropic_Gruneisen_SnSe}, isothermal strains along $a$, $b$, and $c$ axes lead to positive Gr\"{u}neisen parameters i.e., softening of phonon branches along all directions, except for isothermal strain along the $c$ axis, which produces a stiffening of phonon modes along ${\rm \Gamma-X}$ and ${\rm \Gamma-T}$. However, summation over the entire Brillouin zone at constant configuration yields positive effective Gr\"{u}neisen parameters for all cases of isothermal strain as noted in Eq.~\eqref{NTE1} and Table~\ref{Anisotropic_Gruneisen_SnSe}, which is at odds with the experimentally observed negative thermal expansion along the $c$ axis.

\begin{table}
  \caption{Anisotropic Gr\"uneisen parameter ($\gamma$) of tin-selenide calculated by expanding and compressing the lattice along  $a$,  $b$, and $c$ axes. The Brillouin-zone integrated values are averages of $\gamma$ over the entire $41\times41\times41$ Monkhorst-Pack q-point mesh, while the values along specific segments of the Brillouin zone (e.g. $\Gamma-X$) are averages over polarizations only. Further details are provided in the text.}
  \label{Anisotropic_Gruneisen_SnSe}
\begin{center}
  \begin{tabular}{cccc}
  \hline
 \multirow{2}{*}{$\gamma^{avg}_{DFT}$} & \multicolumn{3}{c}{\textbf{Isothermal strain}}\\
  \cline{2-4}
   & $\eta_{11}$ ($a$--axis) & $\eta_{22}$ ($b$--axis) & $\eta_{33}$ ($c$--axis)\\
  \hline
  {BZ integration} & 1.44 & 1.35& 0.64\\
  ${\rm \Gamma-X}$ & 3.26 & 1.34 & -1.01\\
  ${\rm \Gamma-Y}$ & 1.58 & 1.48 & 0.27\\
  ${\rm Y-P}$ & 1.27 & 1.35 & 0.69\\
  ${\rm \Gamma-P}$ & 1.69 & 1.38 & 0.61\\
  ${\rm \Gamma-A}$ & 1.66 & 1.43 & 0.88\\
  ${\rm A-Z}$ & 1.05 & 1.59 & 1.97\\
  ${\rm \Gamma-Z}$ & 1.66 & 1.40 & 0.85\\
  ${\rm \Gamma-T}$ & 1.71 & 1.13 & -0.06\\
\hline
  \end{tabular}
  \end{center}
\end{table}

Alternatively, the relationship between the thermodynamic generalized Gr\"{u}neisen tensor and the thermal expansion tensor, $\alpha_{kl}$, can be expressed in terms of the isothermal bulk modulus $B^T$, volume $V$, and heat capacity at constant configuration $C_{\eta}$, as following: \cite{Wallace1}
\begin{align}\label{NTE2}
\gamma_{ij}^{avg} &= \frac{V}{C_{\eta}}\sum\limits_{kl}B^T_{ijkl}\alpha_{kl}.
\end{align} 
The isothermal bulk modulus tensor, $B_{ijkl}^T$, can be determined either from the phonon group velocities along different crystallographic directions, or from DFT simulations by applying isothermal longitudinal and transverse strains. We have calculated $B_{ijkl}^T$ with both approaches. The independent components of $B_{ijkl}^T$, constrained by the orthorhombic symmetry,  are shown in Table~\ref{Bulk_modulus}. The phonon group velocities at 300\,K were obtained from INS on single crystals, previously reported by our group in Ref.~\cite{Chen2015}. The tensor components not listed in table~\ref{Bulk_modulus} are zero. As can be seen in table~\ref{Bulk_modulus}, the values for $B_{ijkl}^T$ obtained from the two approaches are in fair agreement.

\begin{table}
  \caption{Fourth order isothermal bulk modulus tensor ${B_{ijkl}^T}$ calculated from DFT simulations and experimental inelastic neutron scattering (INS) data. Experimental INS data is at T = 300\,K. More details in text.}
  \label{Bulk_modulus}
\begin{center}
  \begin{tabular}{l|c|c}
  \hline
 ${ B_{ijkl}^T}$ & DFT ({\rm GPa}) & Exp. ({\rm GPa}) \\
  \cline{1-3}
${ B^T_{1111}}$ & 85.5 & 82.5 \\
$ {B^T_{2222}}$ & 89.1& 97.6 \\
${ B^T_{3333}}$ & 52.0 & $\cdots$ \\
${ B^T_{3322} = B^T_{2233}}$ & 41.2 & $\cdots$ \\
${ B^T_{3311} = B^T_{1133}}$ & 23.5 & $\cdots$ \\
${ B^T_{1122} = B^T_{2211}}$ & 19.7 & $\cdots$ \\
${ B^T_{3223} = B^T_{2323} = B^T_{2332} = B^T_{3232}}$  & 49.8 &  31.4\\
${ B^T_{3113} = B^T_{1313} = B^T_{1331} = B^T_{3131}}$  & 22.9 & 17.9 \\
${ B^T_{2112} = B^T_{1212} = B^T_{1221} = B^T_{2121}}$  & 19.1 & 23.7 \\
   \hline
  \end{tabular}
  \end{center}
\end{table}

From the experimentally measured thermal expansion coefficient \cite{Wiedemeier1979}, heat capacity (calculated from experimentally measured phonon DOS, see Fig.~\ref{DOS_SnSe}), volume \cite{Adouby}, and bulk modulus (DFT values were used when experimental values were unavailable) at T = 300\,K, the thermodynamic generalized Gr\"{u}neisen tensor computed from Eq.~\eqref{NTE2} is:
\begin{align}\label{NTE3}
\gamma_{ij, {\rm Exp}}^{\rm avg} &= \begin{bmatrix}     1.03     &    0   &      0 \\
         0  &  0.22    &     0 \\
         0    &     0  & -0.10
         \end{bmatrix}.
\end{align}
Thus, $\gamma_{ij, {\rm Exp}}^{\rm avg}$ (Eq.~\eqref{NTE3}), and $\gamma_{ij, {\rm DFT}}^{\rm avg}$ (Eq.~\eqref{NTE1}) are quite different. 
The Gr\"{u}neisen parameter estimated from experimental data is negative along the $c$ axis, but the DFT prediction is positive. Importantly, while we do observe a substantial negative $\gamma$  along ${\rm \Gamma -X}$ by applying the strain along $c$ (cf Table~\ref{Anisotropic_Gruneisen_SnSe}), this alone is insufficient to produce negative $\gamma$ along $c$ as a whole, once the integration is performed over the entire Brillouin zone. We note that this is not a failure of the DFT calculation \textit{per se}, but rather arises because of the omission of couplings between different crystallographic axes.

The difficulty to capture the NTE along $c$ from the Gr\"{u}neisen parameter prompted us to instead compute the three-dimensional free energy surface, on a 3D grid of isothermal strains. To construct the total 3-D free energy surface, we created a $3\times3\times3$ grid of lattice parameters with 1\% expansion and compression of lattice parameter along each crystallographic axes with total of 27 grid points. The inherent advantage of this approach is that it captures the evolution of both electronic and vibrational free energy by varying lattice parameters individually (i.e., \{$a\pm\Delta a, b, c$\}, \{$a, b\pm\Delta b, c$\}, and so on), as well as coupled terms (i.e., \{$a\pm\Delta a, b\pm\Delta b, c$\}, \{$a\pm\Delta a, b\pm\Delta b, c\pm\Delta c$\}, and so on), and determine the total free energy minima. The total free energy is a sum of electronic free energy, $F_{el}$, and vibrational free energy, $F_{vib}$, which can be expressed as following:
\begin{align}\label{NTE4}
F_{el} &= E_{gs} - TS_{el},
\end{align}
and 
\begin{align}\label{NTE5}
F_{vib} &= \frac{1}{2}\sum\limits_i^{3N}\varepsilon_i + k_{\rm B}T\sum\limits_i^{3N}\ln\,\left\{1-\exp(-\frac{\varepsilon_i}{k_{\rm B}T})\right\},
\end{align}
where electronic entropy, $S_{el}$, and the Fermi-Dirac occupation function, $f$, are given by: 
\begin{align}\label{NTE6}
S_{el} &= -k_{\rm B}\int\limits^{\infty}_{-\infty}[(1-f)\ln(1-f) + f\ln f], 
\end{align}
and 
\begin{align}\label{NTE7}
f &= \frac{1}{\exp\{(E-E_F)/k_{\rm B}T\}+1},
\end{align}
respectively. Here, $E_{gs}$ is ground state energy, $\varepsilon_i$ is phonon energy of $i^{th}$ mode, $E_F$ is the energy at Fermi level, $k_{\rm B}$ is Boltzmann's constant, and $N$ is the number of atoms in the unit cell. We obtain the ground-state electronic energy, and the electronic density of states (eDOS) from DFT simulations for all points on the $(a,b,c)$ grid. The electronic entropy was determined from eq.~\eqref{NTE6}, and~\eqref{NTE7}. The vibrational free energy was obtained from eq.~\eqref{NTE5}, using the phonon density of states (phonon DOS) calculated as described in section~\ref{DFT_simulations}. 

The measured thermal expansion between 300\,K and 813\,K along $a$, $b$, and $c$ is approximately 1.8\%, 3.5\%, and -3.1\%, respectively \cite{Adouby}. Additional points were added to extend our grid to larger expansions/compressions of 3\% along $a$, 5\% along $b$, and 5\% along $c$. However, phonons become unstable for extended grid points. Thus, we limit our procedure to the original 27-point grid, extended with the seven new points for which phonons are stable, shown in Fig.~\ref{ABC_harmonic}(b). To capture the NTE along $c$ over a wider range of temperature, we performed a quadratic extrapolation of the free energy surface (see Eq.~\eqref{NTE8}). We note that the computational cost scales cubicly with the linear grid size, hence the practicality of finer grids is limited.

We determine the equilibrium lattice parameters, $a_{eq}$, $b_{eq}$, and
$c_{eq}$ by minimizing a second-order polynomial fit to the free-energy, $F(a,b,c,T)$, with respect to lattice
parameters at $T=T_0$,
\begin{align}\label{NTE8}
F(a,b,c,T_0) &=  C_{0} + C_1 a + C_2 b + C_3 c + C_4 ab   \nonumber \\
&  + C_5 bc + C_6 ca + C_7 a^2 + C_8 b^2 + C_9 c^2  \nonumber \\
\frac{\partial F(a,b,c,T_0)}{\partial a} &= 0  \nonumber \\
\frac{\partial F(a,b,c,T_0)}{\partial b} &= 0  \nonumber \\
\frac{\partial F(a,b,c,T_0)}{\partial c} &= 0  \; .
\end{align}
The results are shown in Fig.~\ref{ABC_harmonic}(a). As can be seen on this figure, the experimental temperature dependence of lattice parameters is qualitatively captured, including the NTE along $c$. This indicates an interplay between  electronic and vibrational components of the free energy, which ultimately controls the minima of $F$. The deviation between calculated and experimental lattice parameters at high $T$ is partially due to the fact that the theoretical lattice parameters at $T=0$, $a = 11.309, b = 4.119, c = 4.300$\,\AA, are smaller than the experimental values: $a = 11.502, b = 4.153, c = 4.450$\,\AA, leading to shorter theoretical bond lengths, especially in the $b-c$ plane. In addition, strong anharmonic effects, not included in our quasiharmonic calculations of phonon spectra for strained cells, are seen to occur in the last $\sim200$\,K below $T_c$ in our heat capacity measurements (see below). One could also  include anharmonic effects in the phonon calculations, for example with \textit{ab initio} molecular dynamics or other methods incorporating anharmonic phonon renormalization occurring because of the thermal bath \cite{Chen2015, Zhao2014, Carrete2014}. However, our goal is not to exactly match the measured thermal expansion coefficient, but rather to rationalize the origin of the anisotropy and NTE along $c$, which can be explained with this simple and insightful approach. 

\begin{figure}
\begin{center}
\includegraphics[trim=5cm 17.0cm 27cm 0.0cm, clip=true, width=0.8\textwidth]{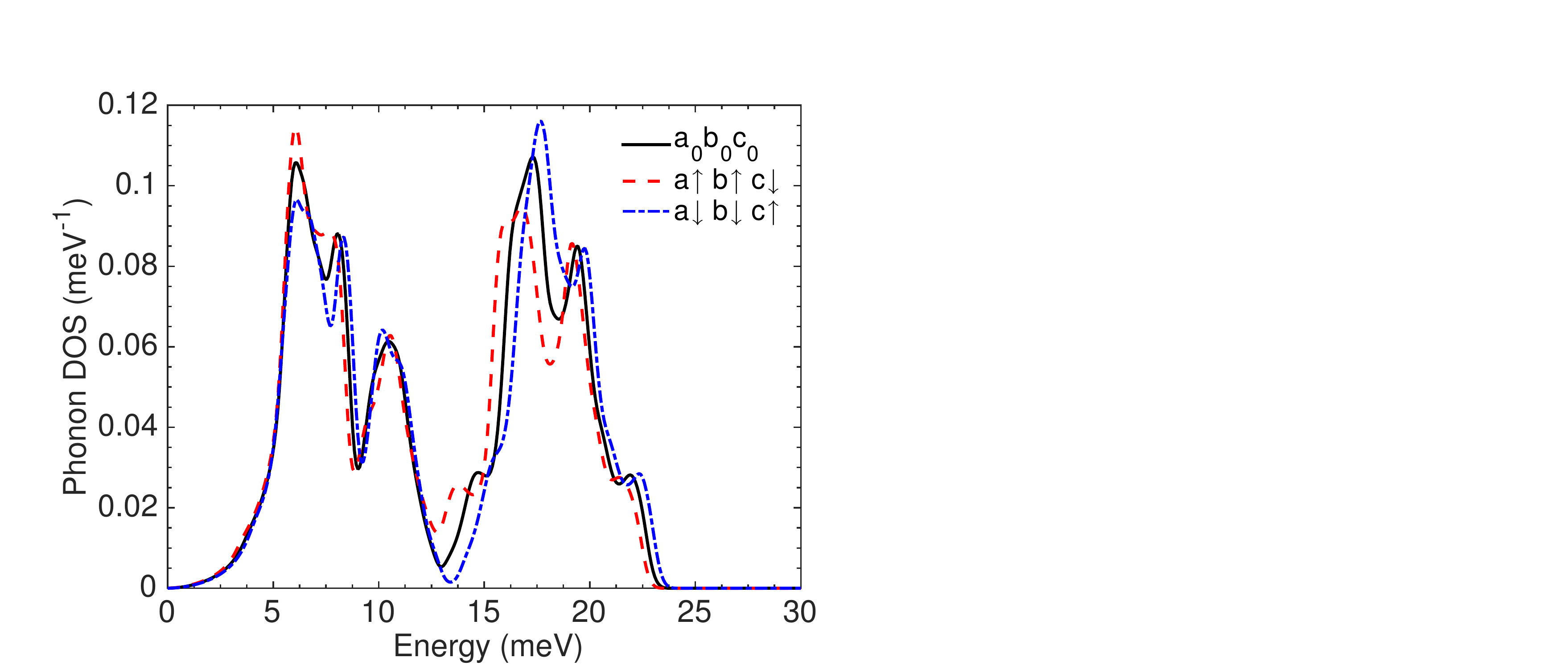}
\end{center}
\caption{\label{DOS_a_b_c} Phonon density of states calculated from DFT for three different configurations of lattice parameters.  The ground-state equilibrium lattice parameters are $a_0b_0c_0$. The other two configurations are $\Delta a/a = \Delta b/b = +1\%, \Delta c/c = -1\%$ $\mathbf{(a\uparrow b\uparrow c\downarrow)}$ and $\Delta a/a = \Delta b/b = -1\%, \Delta c/c = +1\%$ $\mathbf{(a\downarrow b\downarrow c\uparrow)}$.}
\end{figure}

The electronic ground state energy at $T = 0$\,K is minimum for the fully relaxed configuration (relaxed lattice parameters and atomic positions), $(a_0, b_0, c_0)$, and increases with any strain. We note that, for the strained configurations, the minimum in the electronic ground state energy is related to the lattice parameters such that there are two possible scenarios:  $(\Delta a, \Delta b>0, \Delta c<0)$ or  $(\Delta a, \Delta b<0, \Delta c>0)$. Hence, the electronic ground state energy alone can not uniquely define the equilibrium lattice structure. Thus, it is the vibrational free energy that brings the minimum of $F$ towards $(\Delta a, \Delta b>0, \Delta c<0)$. The calculated phonon DOS for three different configurations is shown in Fig.~\ref{DOS_a_b_c}. From this figure, we observe that $(\Delta a, \Delta b>0, \Delta c<0)$ leads to additional spectral weight at low energy in the phonon DOS (``softening''), while $(\Delta a, \Delta b<0, \Delta c>0)$ results in a stiffening. The softening in the first case increases the phonon entropy, and thus lowers $F$. As further discussed below, the configuration $(\Delta a, \Delta b>0, \Delta c<0)$ leads to considerable softening of high energy phonons.

\subsection{Structural Evolution and NTE}

\begin{table}
\caption{Rotation of pyramidal induces NTE along $c$. Bond length and lattice
	constant in $\AA$. All data is from Ref.~\cite{Adouby} except for the calculated
	value $b$ and $c$. The small discrepancy between calculated $c$ and $c_{\rm
	exp}$ results from the ignorance of tilt of two \dtwo\ bonds plane.
	\label{tab:NTE}
}
\begin{center}
	\begin{tabular}{ccccccccc}
	\hline
		& \done & $\theta$ & \dtwo & $\phi$ & $c$ & $b$ & $c_{\rm exp}$ & $b_{\rm exp}$  \\
		\hline
      Cmcm &  2.75  &   0     &     3.05  &   89.85 &   4.32  & 4.31  & 4.31 & 4.31\\ 
      Pnma &  2.72  &  7.71   &     2.80  &   95.67 &   4.49  & 4.15  & 4.45 & 4.15 \\
      \hline
	\end{tabular}
\end{center}
\end{table}

We now investigate the structural changes inside the unit cell in connection with the NTE.
In particular, we show that the NTE of SnSe is closely related to the electronic
instability and its coupling to the anharmonic vibrations of Sn atoms along $c$. 
In a separete work, we showed the strength of \dthr\ bonds is considerably weakened in Pnma compared to Cmcm, as the chemical bond between Sn and Se in \dthr\ essentially breaks \cite{Jiawang-SnSe-arxiv2016}. Thus, the apical \done\ bond and the two in-plane \dtwo\ bonds can be seen to form triangular pyramids with Sn at the apex, as shown in Fig.~\ref{fig:rotation}. 
The phase transition from Cmcm to Pnma is driven by the electronic instability,
which reduces the energy by lifting the degeneracy of four \dtwo\ bonds in-plane
in the Cmcm phase, favoring the off-centering of Sn along $c$, and breaking a mirror plane symmetry.~\cite{Jiawang-SnSe-arxiv2016} 
This distortion of the Cmcm phase overlaps most strongly with a strongly anharmonic soft mode at the zone boundary (optic mode at $Y$ point in Cmcm), coupled to another anharmonic zone-center ($A_g$) optic mode, which together induce rotations of those pyramids as pseudo-rigid units \cite{Jiawang-SnSe-arxiv2016}, and cause an increase in $c$ upon cooling.

In Fig.~\ref{fig:rotation}a, we introduce $\theta$ as the
rotation angle of \done with respect to $a$, and the angle $\phi$ between the \dtwo\ bonds in the
pyramid. For simplicity, we ignore the tilt of the plane formed by two \dtwo\
bonds, away from $b-c$ plane. Within a close approximation, we have:
\begin{align}
	c \simeq 2 d_1 \sin \theta + 2 d_2 \cos \phi/2         
	\label{eq:rotation1}
\end{align}
\begin{align}
        b = 2 d_2 \sin \phi/2  
	\label{eq:rotation2}
\end{align}

\begin{figure}
\includegraphics[width=0.4\textwidth]{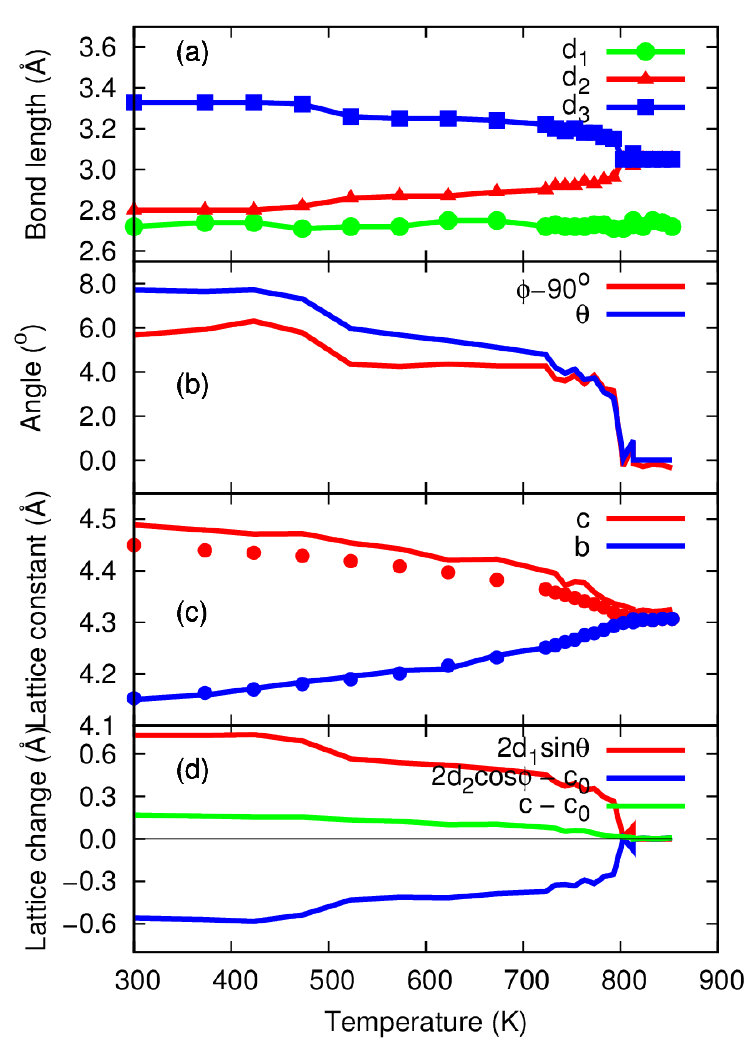}
\caption{\label{fig:bond-angle}
Temperature dependence of bond length (a) and the rotation angle of triangular
pyramids (b) extracted from structure data in Ref.~\citenum{Adouby}. (c) Solids dots is the lattice
parameters in Ref.~\citenum{Adouby}, the curves are derived from Eq.~\ref{eq:rotation1} and~\ref{eq:rotation2}. (d) The $c$ lattice change with temperature, the change is divided by the tilt of $d_1$ bond ($\theta$) and rotation of $d_2$ bond ($\phi$), according to Eq.~\ref{eq:rotation1}, $c_0$ is the lattice constant at 813\,K from data in Ref.~\citenum{Adouby}.}
\end{figure}

From Table~\ref{tab:NTE} we can see that $2 d_2 \cos (\phi/2)$ decreases from
Cmcm to Pnma structure, and therefore, the expansion along $c$ can be seen to result from the
rotation of pyramids ($\theta$), which couples strongly to the $Y + A_g$ mode 
(\done\ almost remains constant).
From the plots in Fig.~\ref{fig:bond-angle}, we can confirm that the NTE along $c$ is indeed
induced by the rotation of pyramids, which is strongly coupled to the anharmonic
modes (soft-modes in Cmcm phase generating zone-center soft-modes in Pnma phase) \cite{Jiawang-SnSe-arxiv2016}. The strongly anharmonic character of these modes partly explains the shortcomings of the quasiharmonic free-energy calculations near the phase transition, seen in Fig.~\ref{ABC_harmonic}.

\section{Anharmonic Phonon Thermodynamics}

\subsection{Phonon DOS and Phonon Entropy}

\begin{figure}
\begin{center}
\includegraphics[trim=2cm 0.0cm 32cm 0.0cm, clip=true, width=0.5\textwidth]{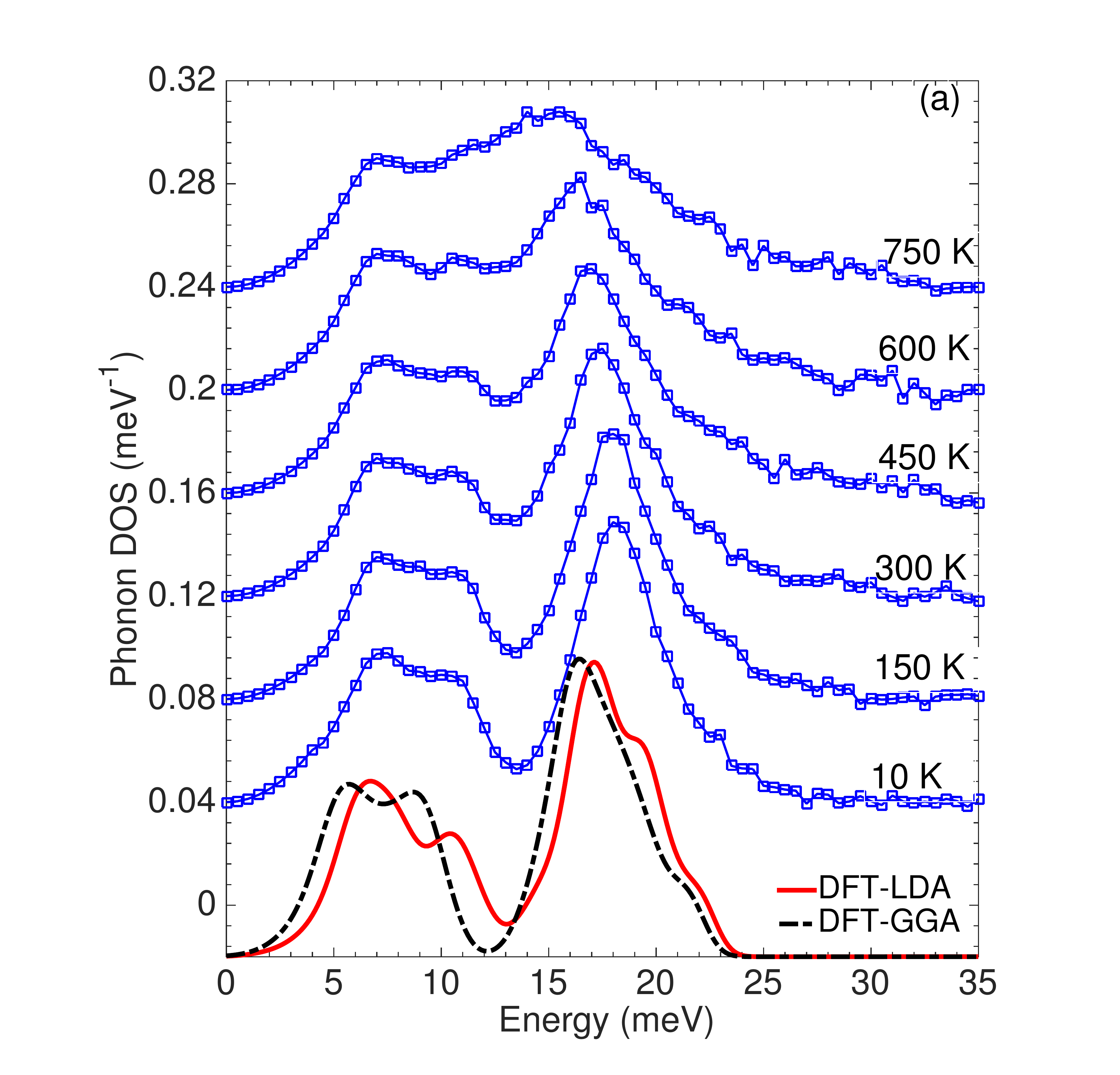}
\includegraphics[trim=2cm 0.0cm 32cm 0.0cm, clip=true, width=0.5\textwidth]{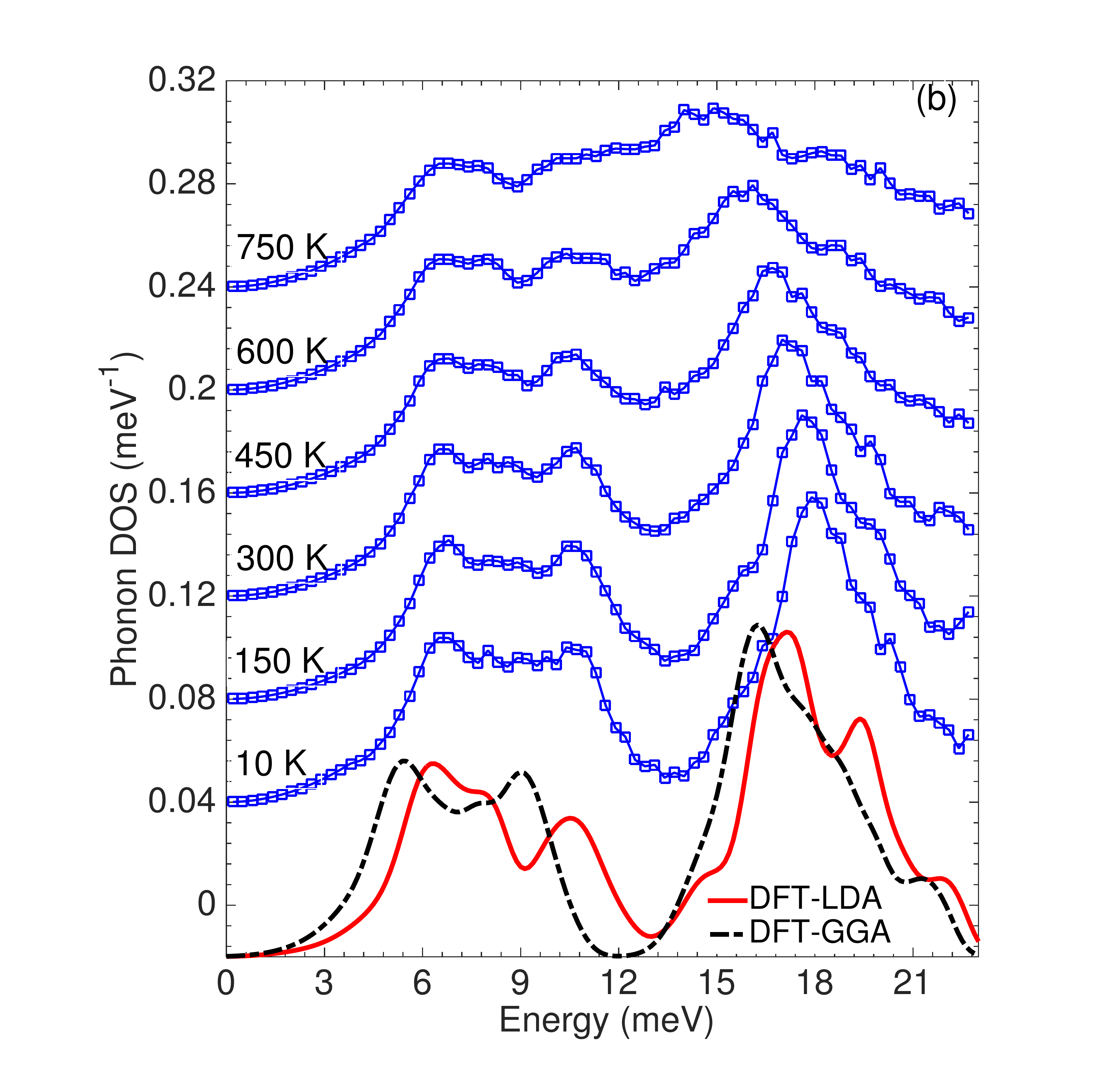}
\end{center}
\caption{\label{DOS_SnSe} Neutron-weighted phonon DOS of $\rm{SnSe}$ measured with inelastic neutron scattering at different temperatures for incident neutron energies of a) 55\,meV and b) 30\,meV, compared with neutron weighted and experimental resolution convoluted DFT simulations.}
\end{figure}

Much of the free-energy discussion above was limited to a quasiharmonic approximation. In order to estimate the anharmonic component of the vibrational entropy, we investigate the temperature dependence of the measured phonon DOS. Fig.~\ref{DOS_SnSe}a and~\ref{DOS_SnSe}b show the experimental phonon DOS measured at different temperatures for $E_i=55$\,meV and 30\,meV, respectively. The low and high energy phonon peaks soften by $\sim$0.3 and $\sim$1.0\,meV, respectively as the sample temperature is increased from 10 to 750\,K. One can observe a drastic broadening of the spectrum with increasing $T$, in particular for optical modes above 13\,meV. While the lowest energy zone-center TO mode has been shown to soften strongly with increasing $T$ in Pnma in our previous INS dispersion measurements (Ref.~\citenum{Chen2015}), that mode is confined to the zone-center and has a limited spectral weight in the DOS. We note that the neutron-weighted phonon DOS from harmonic DFT simulations agrees well with the total and direction-projected phonon DOS measured at 300\,K (Fig.~\ref{partial_DOS_SnSe} and~\ref{DOS_SnSe}). The direction-projected experimental phonon DOS measurements (Fig.~\ref{partial_DOS_SnSe}) provide substantial information in separating the two low energy Sn dominated phonon peaks at $\sim$6 and $\sim$11\,meV. The low energy ($\sim$6\,meV) phonon peak include both out-of-plane (along $a$-axis) and in-plane (b-c plane) Sn vibrations, while $\sim$11\,meV phonon peak primarily originates from in-plane (b-c plane) Sn vibrations. Both Sn dominated phonon peaks show substantial broadening in the phonon spectrum with increasing temperature (Fig.~\ref{DOS_SnSe}). 

With increasing temperature, it is the coupling between lattice distortion (especially the large Sn in-plane displacements) and the electronic structure that generates strong  anharmonicity (leading to phonon-phonon interactions) \cite{Chen2015}, primarily responsible for a significant phonon softening of high energy phonons and phonon broadening of entire phonon spectra seen in Fig.~\ref{DOS_SnSe}. This conclusion is further supported by our DFT simulations of NTE modeling where we identify that the minimization of electronic free energy alone is not sufficient to obtain the experimental lattice parameters with temperature, and including the change in vibrational free energy due to lattice distortions is necessary to reach equilibrium. As we have detailed in Ref.~\citenum{Jiawang-SnSe-arxiv2016}, the origin of the anharmonicity for Sn in-plane motions is the Jahn-Teller instability involving the Se \textit{p}-states.

\begin{figure}
\begin{center}
\includegraphics[trim=2cm 0.0cm 26cm 0.0cm, clip=true, width=0.4\textwidth]{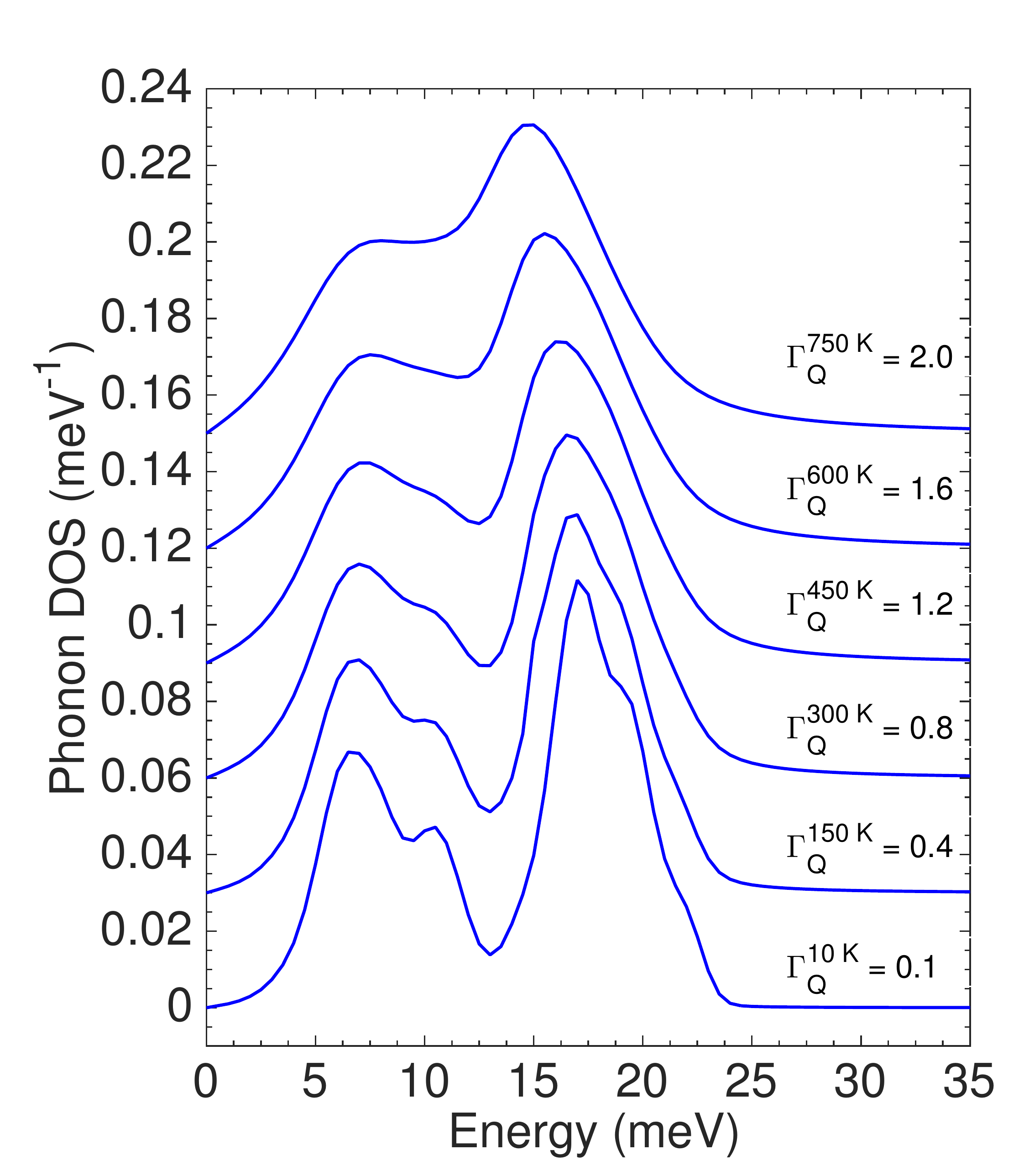}
\end{center}
\caption{\label{NW_DOS_SnSe} Generalized phonon DOS of $\rm{SnSe}$ calculated with DFT, weighted with experimental neutron cross-sections. The phonon DOS is convolved with a damped harmonic oscillator function, as detailed in the text.}
\end{figure}

To quantify the anharmonic contribution to $S_{\rm vib}$, we estimated a phonon broadening parameter, $\Gamma_Q^T$, by convolving the generalized (neutron-weighted) phonon DOS from DFT, $g(E)$, with a damped harmonic oscillator function, $B(E,E',T)$, \cite{Lovesey1984}
\begin{align}\label{NTE9}
g(E,T) = \int B(E,E',T)g(E'-\Delta E'(T))\,dE',
\end{align}
 and matching the result with the measured phonon DOS. Here,
\begin{align}\label{NTE10}
B(E,E',T) = \frac{2\Gamma_Q^T}{(\pi EE')\left[\left\{\frac{E}{E'}-\frac{E'}{E}\right\}^2+\left\{\frac{2\Gamma_Q^T}{E'}\right\}^2\right]},
\end{align}
and $\Delta E'$ is a shift in phonon energy calculated by fitting the Lorentzian to the peaks in the measured phonon DOS. The resulting phonon DOS, $g(E,T)$, with temperature-dependent phonon broadening parameter, $\Gamma_Q^T$, is shown in Fig.~\ref{NW_DOS_SnSe}. This model reproduces the temperature dependence of the experimental phonon DOS well, and enables us to assess the anharmonic $S_{\rm vib}$. For this purpose, we use the {\it non-neutron-weigthed} version of $g(E,T)$.

The vibrational entropy $S_{vib}$ can be expressed as \cite{Fultz2010,Wallace1,Wallace2},
\begin{align}\label{NTE12}
S_{vib} &= 3N\,k_{\rm B} \nonumber \\
&\times\int g(E,T)\left[(1+n_T)\ln(1+n_T) - n_T\ln(n_T)\right]~dE \; ,
\end{align}
where $n_T(E) = [e^{E/k_{\rm B}T}-1]^{-1}$ is the Bose-Einstein occupation factor for phonons, and $N$ is the number of atoms in the crystal.

The harmonic phonon entropy, $S_{vib,h}$, was calculated by convolving the harmonic phonon DOS from DFT with a damped anharmonic oscillator function $B(E,E',T)$ at $T=10$\,K. The resulting phonon DOS was substituted in Eq.~\eqref{NTE12} to evaluate $S_{vib,h}$ at different temperatures. To evaluate the total entropy, $S_{vib}$, at different temperatures, the $T$--dependent damped anharmonic oscillator function $B(E,E',T)$ was convolved with the DFT phonon DOS, and substituted in Eq.~\eqref{NTE12}. The vibrational entropy due to thermal expansion of the lattice is given by
\begin{align}\label{NTE13}
S_{vib,d} &= \int\limits_0^T~\frac{(C_p-C_v)}{T'}~dT'  \nonumber \\
&= \int \limits_0^T~\frac{\alpha_{vib}(T')^2V(T')}{\chi(T')}~dT',
\end{align}
where $\alpha_{vib}$ is the volumetric thermal expansion coefficient, $\chi$ is the compressibility (inverse of bulk modulus), and $C_p$ and $C_v$ are specific heat capacity of material at constant pressure and temperature. 

\begin{figure}
\begin{center}
\includegraphics[trim=29.0cm 0.0cm 4cm 0.0cm, clip=true, width=0.45\textwidth]{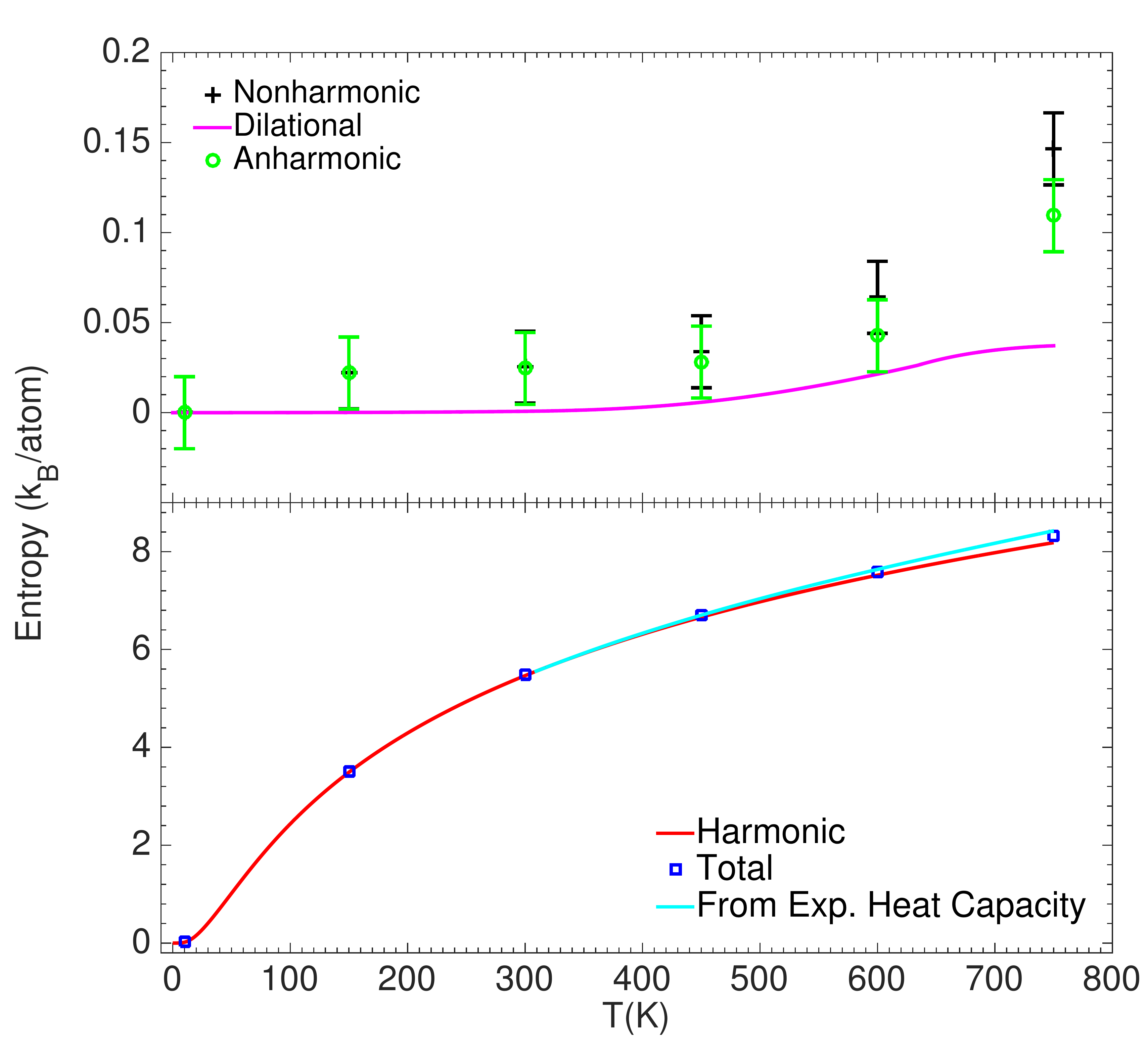}
\end{center}
\caption{\label{entropy} Components of vibrational entropy, $S = S(T) - S(T=0\,K)$, of $\mathrm{SnSe}$ at T = 10, 150, 300, 450, 600, and 750\,K. Lower panel shows the markers of total ($\square$) and harmonic (red line) vibrational entropy contribution along with values calculated from experimental heat capacity data (cyan line), while upper panel displays the markers for nonharmonic (+), dilational (magenta line), and anharmonic entropy ($\circ$). More details in the text.}
\end{figure}

Using the total $S_{vib}$, harmonic $S_{vib,h}$, and dilational $S_{vib,d}$ components of the vibrational entropy, the nonharmonic and anharmonic entropy were calculated as: $S_{vib,nh} = S_{vib} - S_{vib,h}$, and $S_{vib,ah} = S_{vib} - S_{vib,h} - S_{vib,d}$, respectively. For the calculation of dilational entropy, we used the thermal expansion coefficient from Wiedemeier $\emph{et al.}~$\cite{Wiedemeier1979}, and temperature dependent compressibility from He $\emph{et al.}$~\cite{He2013}. In addition, we have also  estimated the total entropy from our experimental heat capacity measurements ($S_{\rm DSC}$) using the following expression,
\begin{align}\label{NTE13_2}
S_{\rm DSC} = \int\limits_{T_0}^{T}\frac{C_p}{T}\,dT.
\end{align}
The calculated harmonic, dilational, non-harmonic, anharmonic, and total vibrational entropy values are shown in Fig.~\ref{entropy}. 

\begin{figure}
\begin{center}
\includegraphics[trim=29.0cm 0.0cm 4cm 0.0cm, clip=true, width=0.45\textwidth]{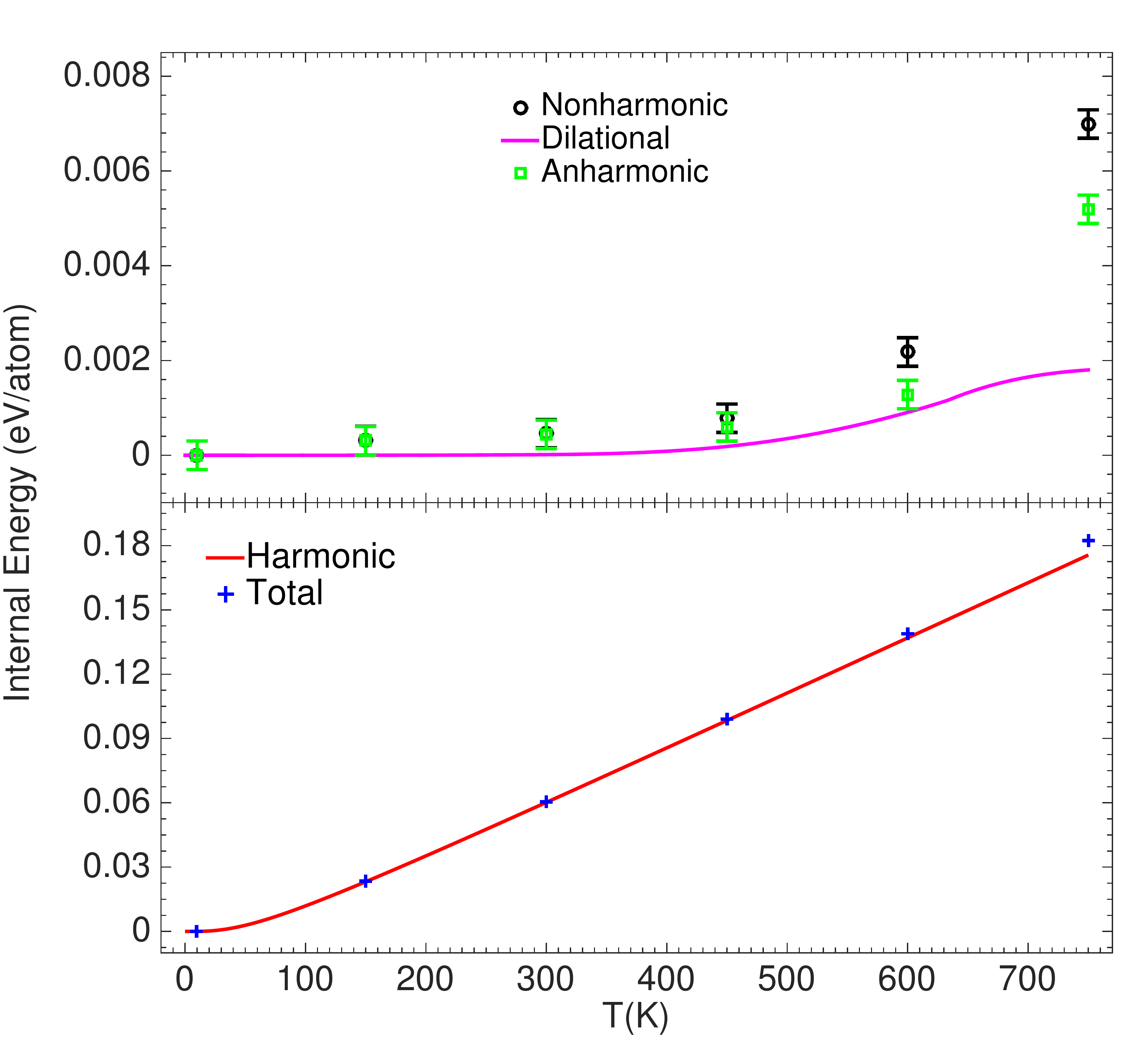}
\end{center}
\caption{\label{internal_energy} Components of vibrational internal energy, $E = E(T) - E(T=0\,K)$, of $\mathrm{SnSe}$ at T = 10, 150, 300, 450, 600, and 750\,K. Lower panel shows the markers of total ($\square$) and harmonic (red line) vibrational internal energy contribution, while upper panel displays the markers for nonharmonic (+), dilational (magenta line), and anharmonic internal energy ($\circ$). More details in the text.}
\end{figure}

\begin{figure}
\begin{center}
\includegraphics[trim=29.0cm 0.0cm 4cm 0.0cm, clip=true, width=0.45\textwidth]{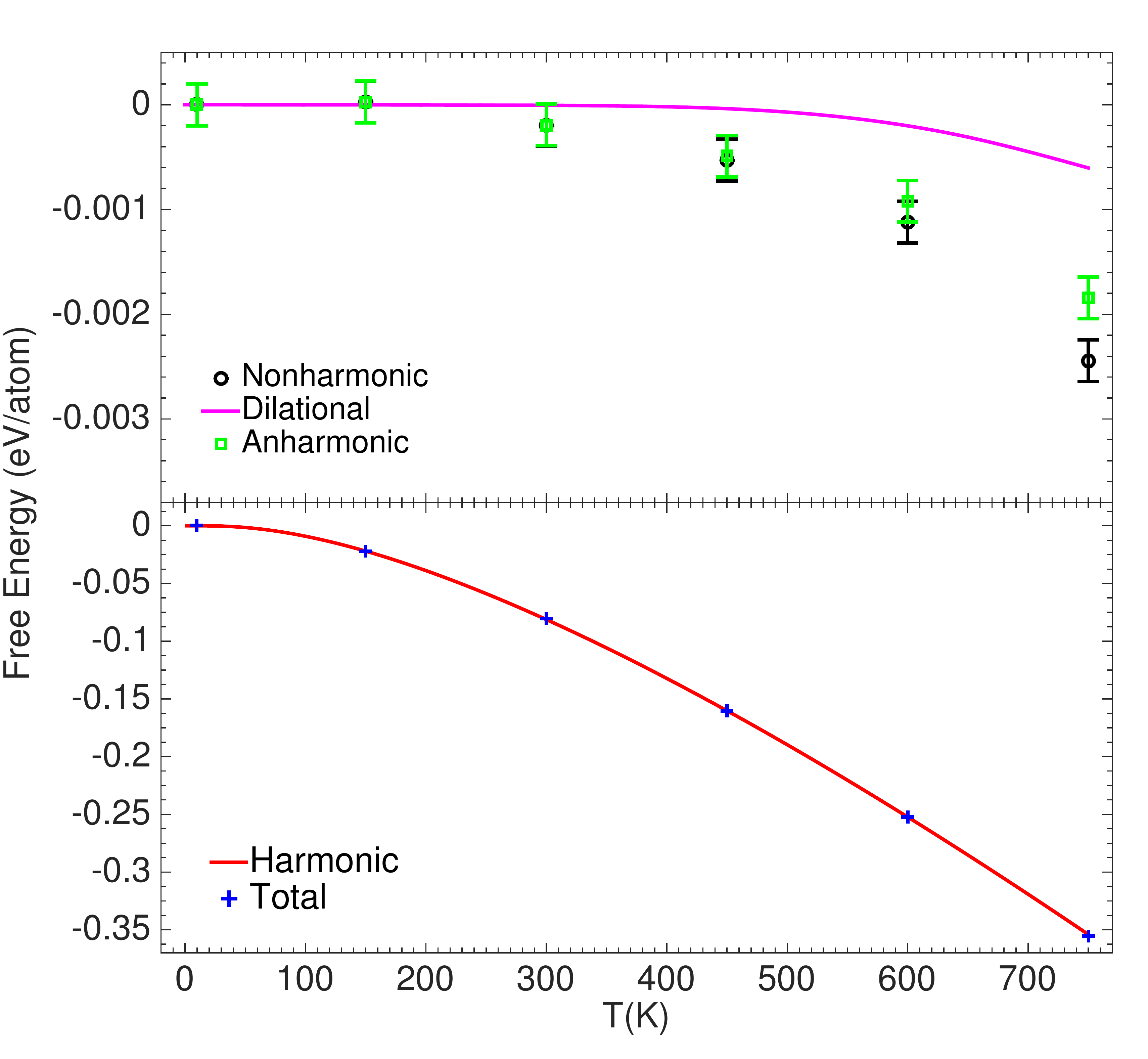}
\end{center}
\caption{\label{free_energy} Components of vibrational free energy, $F = F(T) - F(T=0\,K)$, of $\mathrm{SnSe}$ at T = 10, 150, 300, 450, 600, and 750\,K. Lower panel shows the markers of total ($\square$) and harmonic (red line) vibrational free energy contribution, while upper panel displays the markers for nonharmonic (+), dilational (magenta line), and anharmonic free energy ($\circ$). More details in the text.}
\end{figure}

As we can observe from Fig.~\ref{entropy}, the main entropic contribution is the harmonic term. However, as temperature increases, the nonharmonic contribution becomes significant and cannot be ignored. The nonharmonic entropy is a summation of dilational and anharmonic entropy contributions. We have also calculated nonharmonic contributions to the vibrational internal energy and vibrational free energy following methods recently reported in Ref.~\citenum{Bansal_2016_1}. We should note that since tin-selenide is a indirect band gap semiconductor (band gap of 0.86\,eV in Pnma phase at room temperature)\cite{Zhao2014}, we could expect the contribution of electronic excitations to specific heat capacity/thermal expansion coefficient to be negligible at room temperature. The band gap in SnSe is temperature dependent, and reduces to $\sim 0.4$\,eV in Cmcm phase \cite{Zhao2014}. While the temperature dependence of band gap may affect our results slightly, we do not expect this effect to be significant. From these calculations, a similar trend is observed in vibrational internal energy (Fig.~\ref{internal_energy}) and vibrational free energy (Fig.~\ref{free_energy}), with the anharmonic contribution rising as $T_c$ is approached. As we can observe, at high $T$, thermal expansion (dilation) alone cannot account for nonharmonic contributions, and anharmonicity plays a large role near the instability.

It is interesting to note that in the experimental phonon DOS (Fig.~\ref{DOS_SnSe}), there is considerably larger softening, ${\Delta E}/{\langle E\rangle} \sim-5.2$\%, of high energy phonons ($E\ge13$\,meV) compared to the softening of low energy phonons, $\sim-3.8$\%, between $T=10$ and 750\,K. As evident from Fig.~\ref{partial_DOS_SnSe}a, low energy and high energy modes are dominated by ${\rm Sn}$ and ${\rm Se}$, respectively. We selectively investigate the change in low and high energy phonon energies by applying 1\% isothermal strain along $\Delta a/a = \Delta b/b = +1\%, \Delta c/c = -1\%$ (3-D free energy surface minima direction as described earlier), and find that high energy phonons soften by $\sim-2.1$\% in comparison to $\sim-1.0$\% for low energy phonons. 

The values of Gr\"uneisen parameter along high-symmetry directions are reported in Table~\ref{Isothermal_Gruneisen_SnSe}. The large values of Gr\"uneisen parameter across the various high symmetry directions in ${\rm SnSe}$ reflect the pronounced anharmonicity, which is also shown quantitatively in the anharmonic vibrational entropy, internal energy and free energy calculations, while the significant variation in values across different directions is attributed to the structural anisotropy in this material.
\begin{table}
  \caption{Gr\"uneisen parameter ($\gamma$) of tin-selenide calculated by applying 1\% isothermal strain along $\Delta a/a = \Delta b/b = +1\%, \Delta c/c = -1\%$ $\mathbf{(a\uparrow b\uparrow c\downarrow)}$. Average along particular direction such as ${\rm \Gamma-X}$ represents the values of $\gamma$ averaged over q-point along that direction weighted by specific heat capacity. More details in text.}
  \label{Isothermal_Gruneisen_SnSe}
\begin{center}
  \begin{tabular}{c|c|c|c}
  \hline
 \multirow{2}{*}{$\gamma^{avg}_{DFT}$} & \multicolumn{3}{c}{\textbf{Isothermal strain along: $\mathbf{(a\uparrow b\uparrow c\downarrow)}$}}\\
  \cline{2-4}
   & All modes & modes $<$ 13\,meV & modes $\ge$ 13\,meV\\
  \hline
  ${\rm \Gamma-X}$ & 5.51 & 6.49 & 4.49\\
  ${\rm \Gamma-Y}$ & 2.79 & 2.92 & 2.66\\
  ${\rm Y-P}$ & 1.91 & 1.39 & 2.46\\
  ${\rm \Gamma-P}$ & 2.46 & 2.08 & 2.86\\
  ${\rm \Gamma-A}$ & 2.20 & 1.79 & 2.63\\
  ${\rm A-Z}$ & 0.63 & -0.39 & 1.68\\
  ${\rm \Gamma-Z}$ & 2.21 & 1.83 & 2.61\\
  ${\rm \Gamma-T}$ & 2.90 & 2.59 & 3.23\\
\hline
  \end{tabular}
  \end{center}
\end{table}

\subsection{Partial Phonon DOS and Thermal Displacement Parameters}

\begin{figure*}
\begin{center}
\includegraphics[trim=5cm 17.0cm 5cm 0.0cm, clip=true, width=1.0\textwidth]{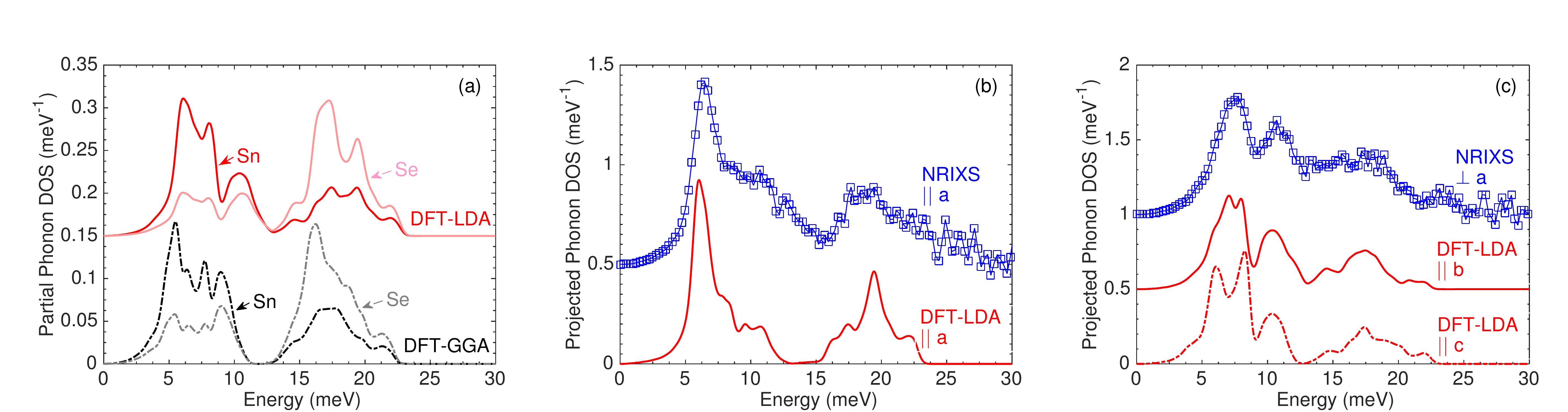}
\end{center}
\caption{\label{partial_DOS_SnSe} a) Partial phonon DOS of tin and selenium calculated from DFT simulations for $a_0b_0c_0$, the ground state equilibrium lattice parameters for fully relaxed structure, b) projected phonon DOS of Sn parallel to $a$ axis, and c) projected phonon DOS of Sn parallel to $b$ and $c$ axis compared with partial phonon DOS of Sn from NRIXS measurements at room temperature.}
\end{figure*}

Fig~\ref{partial_DOS_SnSe}-a shows the partial phonon DOS of tin and selenium calculated from DFT simulations for $a_0b_0c_0$, the ground state equilibrium lattice parameters for fully relaxed structure with Sn and Se atoms predominantly occupying the low and high energy phonon spectrum, respectively. To enable the direct comparison with experimentally measured NRIXS Sn partial phonon DOS parallel and perpendicular to $a$-axis, we have also calculated the projected phonon DOS of Sn as shown in Fig~\ref{partial_DOS_SnSe}-b, and~\ref{partial_DOS_SnSe}-c. The good agreement between experiment and simulations further validates the accuracy of DFT-LDA. Additionally, from the partial phonon DOS and f-factor (Lamb-M\"ossbauer factor) measured from NRIXS \cite{Sturhahn1999, Sturhahn2000}, the mean square thermal displacement of Sn can be calculated. The f-factors corresponding to partial phonon DOS parallel and perpendicular to $a$-axis are 0.1370$\pm$0.0098 and 0.1487$\pm$0.0095, respectively. We should note that, the exact sample orientation for in-plane ($\perp$ a) was not known; however, the in-plane NRIXS measurement agrees well with the simulation for the $b$-axis and we believe this was the likely orientation. 

Furthermore, we have calculated the mean square thermal displacement parameter, $\langle|U_{\alpha}(j)|^2\rangle$, from DFT phonon spectra, as:
\begin{align}
\langle|U_{\alpha}(j)|^2\rangle = \frac{\hbar}{2Nm_j}\sum\limits_{{\bm q},s}\omega_s({\bm q})^{-1}(1+2n_{s}({\bm q}))|e_{s,{\alpha}}(j,{\bm q})|^2,
\end{align} 
where, $\alpha$ denotes thermal displacement direction, $m_j$ is mass of atom at location $j$, $\omega_s({\bm q})$ is phonon frequency of $s^{th}$ phonon branch at wave vector  ${\bm q}$, $n_s$ is the mean Bose-Einstein occupation factor, and $e_{s,{\alpha}}(j,{\bm q})$ is phonon wave vector.  The results for $T=300\,$K, listed in Table~\ref{Thermal_displacement_SnSe}, are consistent with values measured by Chattopadhyay $\emph{et al.}$ at room temperature with neutron diffraction \cite{Chatto1986}. The thermal displacement parameters for ${\rm Sn}$ and ${\rm Se}$ have similar magnitudes at 300\,K, thus indicating that the average bond stiffness at each site is comparable. Indeed, our DFT simulations give a small increase ( $\sim$12\%) in the trace of Se on-site force-constant matrix in comparison to Sn. The ratio of average frequency for Se and Sn vibrations calculated from DFT is 1.38, close to the effect of mass ratio, $\sqrt{{M(Sn)}/{M(Se)}} \approx 1.23$.  It is worth emphasizing, that while the magnitude of thermal displacements of Se atoms reported by Chattopadhyay $\emph{et al.}$ increases linearly with temperature, off-centered Sn atoms show a superlinear increase for in-plane thermal displacements near the phase transition \cite{Chatto1986, Adouby}, which reflects the strong anharmonicity in this regime.

\begin{table*}
  \caption{Mean square thermal displacement parameter $\langle|U_{\alpha}(j)|^2\rangle$ of tin and selenium calculated from DFT (LDA) simulations at T = 300\,K compared with experimental neutron diffraction and NRIXS data. }
  \label{Thermal_displacement_SnSe}
\begin{center}
  \begin{tabular}{|c|c|c|c|c|c|}
  \hline
$\langle|U_{\alpha}(j)|^2\rangle$  & \multicolumn{3}{c|}{${\rm Sn}$}& \multicolumn{2}{c|}{${\rm Se}$}\\
  \cline{2-6}
   (in \AA$^2$) & This work (DFT) & This work (NRIXS) & Ref.~\cite{Chatto1986} & This work (DFT) & Ref.~\cite{Chatto1986}\\
  \hline
  ${\rm U_{11}}$ & 0.0165 & 0.0136 & 0.0151 & 0.0143 & 0.0130\\
  ${\rm U_{22}}$ & 0.0148 & 0.0130 & 0.0130 & 0.0118 & 0.0107\\
  ${\rm U_{33}}$ & 0.0170 & -- & 0.0177 & 0.0126 & 0.0143\\
\hline
  \end{tabular}
  \end{center}
\end{table*}

\subsection{Heat capacity}

\begin{figure}
\begin{center}
\includegraphics[trim=4.5cm 13.8cm 17cm 1.0cm, clip=true, width=0.9\textwidth]{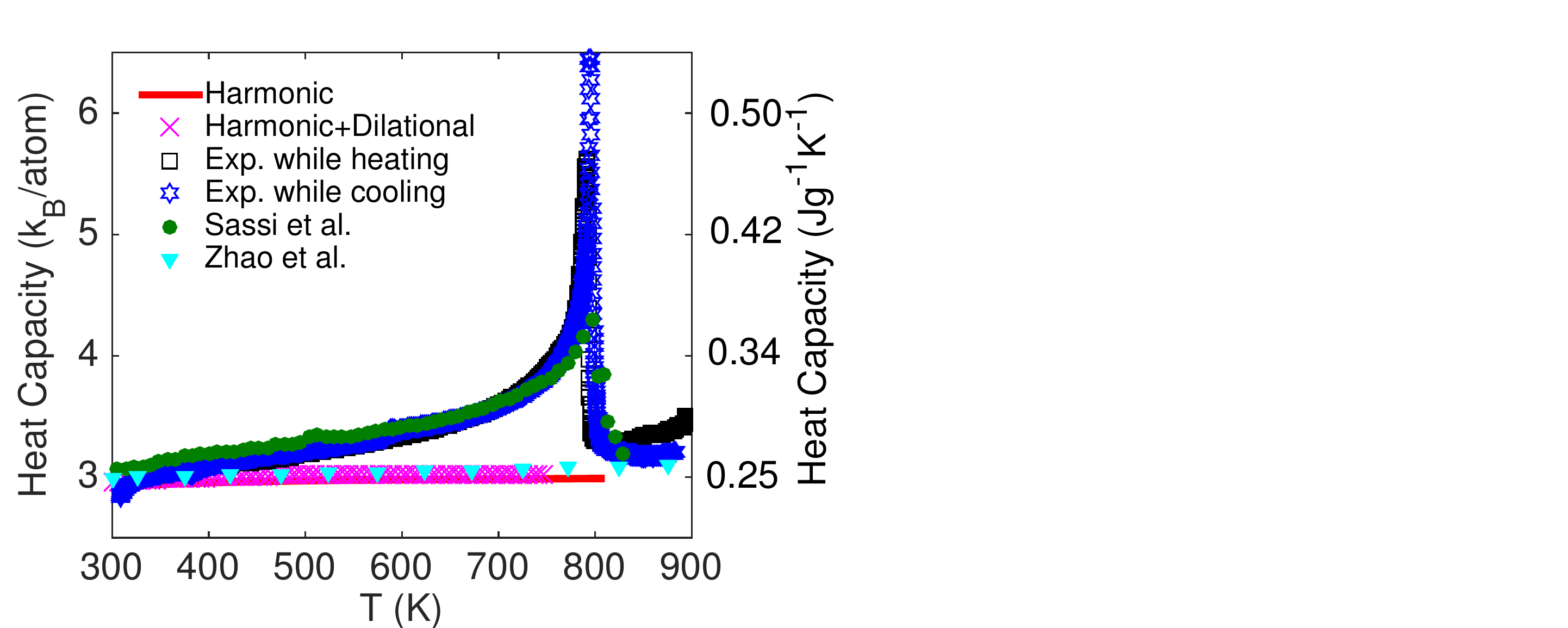}
\end{center}
\caption{\label{heat_capacity} Heat capacity of SnSe measured with DSC on single-crystal samples (both heating and cooling), and its harmonic, dilational, and nonharmonic components, indicating the strong anharmonicity around the phase transition. The green dots are data measured on polycrystalline samples by Sassi {\it et al.} \cite{Sassi2014}. The cyan triangles are data measured on single crystal samples by Zhao {\it et al.} \cite{Zhao2014}}
\end{figure}

To further investigate the nature of the phase transition and anharmonic effects in SnSe, we measured the heat capacity. The measurements were performed with a Netzsch DSC 404C differential scanning calorimeter, with the sample loaded inside a Pt crucible, under an ultra-pure Ar purge gas cycled through a Ti gettering furnace. The scans were performed after careful evacuation and purging of the sample chamber. The heating and cooling rates were 20\,K/min and 20\,K/min, respectively. A sapphire standard and empty-crucible baseline measurements were performed in identical conditions. The heat capacity curves measured during heating and cooling are shown in Fig.~\ref{heat_capacity}. The heat capacity exhibits a lambda shape in the vicinity of the phase transition, akin to the classic case of liquid helium \cite{Buckingham1961}. This behavior is generally indicative of a second-order phase transition, in agreement with the nearly continuous evolution of structural parameters observed with diffraction \cite{Adouby, Wiedemeier1979, Chatto1986}. The $c$-polarized lowest-energy transverse optic soft-mode was also shown to continuously condense across $T_c$ in our previous INS study  \cite{Chen2015}. The phase transition temperature, $T_c\sim795\pm4$\,K,  obtained from our heat capacity measurements is in good agreement with values (802--813\,K) reported in the literature \cite{Adouby, Wiedemeier1979, Chatto1986}. A slight hysteresis in $T_c$ of about 4\,K was observed in our DSC. Some of the hysteresis may possibly be caused by some Se evaporation at high $T$, oxidation, or temperature lags in the instrument. However, it is also possible for the transition to exhibit a partially first-order character, since there is some reported evidence for a small latent heat \cite{Feutelais1996, Sharma1986, Balde1981}. We point out that the values of $C_p$ from our DSC measurements (Fig.~\ref{heat_capacity}), while in excellent agreement with those of Sassi {\it et al.} \cite{Sassi2014}, are significantly larger than the linear estimate from laser flash measurements reported in Ref.~\citenum{Zhao2014}, which also misses the phase transition behavior. For example, at 700\,K, our $C_p$ measurement is 16\% larger than the linear estimate of Zhao {\it et al.} \cite{Zhao2014}, and at 790\,K the discrepancy approximately reaches a factor of two. However, the discrepancy is more minimal above the phase transition, in the Cmcm phase.

We calculated the harmonic and dilational heat capacity from INS phonon DOS and thermal expansion coefficient measurements. The harmonic ($C_v$)and dilational ($C_d$) heat capacity are given by,
\begin{align}\label{NTE14}
C_v = k_{\rm B}\left(\frac{\epsilon_i}{k_{\rm B}T}\right)\frac{\exp(\epsilon_i/k_{\rm B}T)}{(\exp(\epsilon_i/k_{\rm B}T)-1)^2}, \text{ and}
\end{align}
\begin{align}\label{NTE15}
C_d = C_p - C_v = \frac{\alpha_{vib}(T')^2V(T)T}{\chi(T)},
\end{align}
respectively \cite{Wallace1,Fultz2010}. To calculate the harmonic heat capacity from Eq.~\eqref{NTE14}, we have used the non-neutron-weigthed version of phonon DOS at 10\,K, as described earlier for the entropy calculations. The comparison of experimental measurements with harmonic and dilational heat capacity is shown in Fig.~\ref{heat_capacity}. For $T<500$\,K, the quasi-harmonic (harmonic+dilational) heat capacity accounts well for the measured $C_p$. However, as the phase transition is approached, the difference between the two curves increases significantly, reflecting the growing contribution of anharmonicity. A similar behavior was also seen in entropy, internal energy and free energy (Fig.~\ref{entropy},~\ref{internal_energy}, and~\ref{free_energy}). 

\section{Conclusion}

We have investigated the anharmonic thermodynamics and negative thermal expansion in SnSe with a combination of INS, NRIXS, calorimetry measurements, and first-principles simulations. We identified a pronounced contribution of anharmonicity at high temperature, especially within $\sim 200$\,K of the structural phase transition. The NTE along the $c$--axis, the direction of corrugation of Sn-Se bilayers in the Pnma phase, can be qualitatively accounted for with a quasiharmonic free energy minimization, but deviations arising from strong anharmonicity are evident at high temperatures. The structural distortion accompanying the NTE can be traced to rotations of SnSe$_3$ tetrahedra, which overlap with the strongly anharmonic soft-modes condensing at the phase transition.  

\section{Acknowledgements}
We would like to acknowledge Ayman H. Said for technical help with NRIXS measurements at APS sector 30. We also thank Amr Mohammed for help in preparing DSC samples and measurements. Data analysis, modeling, and phonon simulations (D.B, J.H.) were supported by the U.S. Department of Energy, Office of Science, Basic Energy Sciences, Materials Sciences and Engineering Division, through the Office of Science Early Career Award grant of O.D. Neutron scattering measurements were supported as part of the S3TEC EFRC, an Energy Frontier Research Center funded by the US Department of Energy, Office of Science, Basic Energy Sciences under Award \# DE-SC0001299 (C.W.L, O.D.). Sample synthesis (A.F.M.) was supported by the US Department of Energy, Office of Science, Basic Energy Sciences, Materials Sciences and Engineering Division. The use of Oak Ridge National Laboratory's Spallation Neutron Source was sponsored by the Scientific User Facilities Division, Office of Basic Energy Sciences, US Department of Energy. Use of the APS was supported by DOE-BES under Contract No. DE-AC02-06CH11357. Theoretical calculations were performed using resources of the National Energy Research Scientific Computing Center, a DOE Office of Science User Facility supported by the Office of Science of the US Department of Energy under contract no.DE-AC02-05CH11231. This research used resources of the Oak Ridge Leadership Computing Facility, which is supported by the Office of Science of the U.S. DOE.


%

\end{document}